\documentclass[%
reprint,
amsmath,amssymb,
 aps,
prl,
superscriptaddress
]{revtex4-1}

\usepackage{graphicx}
\usepackage{dcolumn}
\usepackage{bm}
\usepackage{stmaryrd}
\usepackage{pxfonts}
\usepackage{txfonts}
\usepackage{float}
\usepackage{ulem}
\usepackage{amssymb}
\usepackage{amsmath}
\usepackage[dvipsnames]{xcolor}
\usepackage{textcomp}
\usepackage{titlesec}
\usepackage{color}
\usepackage{multirow}
\usepackage{afterpage}

\begin{document}

\title{Detection of relativistic fermions in Weyl semimetal TaAs by magnetostriction measurements}

\author{T.\,Cichorek}
\affiliation{Institute of Low Temperature and Structure Research, Polish Academy of  Sciences, 50-422 Wroc{\l}aw, Poland}
\author{L.\,Bochenek}
\affiliation{Institute of Low Temperature and Structure Research, Polish Academy of  Sciences, 50-422 Wroc{\l}aw, Poland}
\author{J.\,Juraszek}
\affiliation{Institute of Low Temperature and Structure Research, Polish Academy of  Sciences, 50-422 Wroc{\l}aw, Poland}
\author{Yu.\,V.\,Sharlai}
\affiliation{B. Verkin Institute for Low Temperature Physics and Engineering, Ukrainian Academy of Sciences, Kharkov 61103, Ukraine}
\author{G.\,P.\,Mikitik}
\affiliation{B. Verkin Institute for Low Temperature Physics and Engineering, Ukrainian Academy of Sciences, Kharkov 61103, Ukraine}

\begin{abstract}
Thus far, a detection of the Dirac or Weyl fermions in topological semimetals remains often elusive, since in these materials conventional charge carriers exist as well. Here, measuring a field-induced length change of the prototype Weyl semimetal TaAs at low temperatures, we find that its $c$-axis magnetostriction amounts to relatively large values whereas the $a$-axis magnetostriction exhibits strong variations with changing the orientation of the applied magnetic field. It is discovered that at magnetic fields above the ultra-quantum limit, the magnetostriction of TaAs  contains a linear-in-field term, which, as we show, is a hallmark of the Weyl fermions in a material. Developing a theory for the magnetostriction of noncentrosymmetric topological semimetals and applying it to TaAs, we additionally find several  parameters characterizing the interaction between the relativistic fermions and elastic degrees of freedom in this semimetal. Our study shows how dilatometry can be used to unveil Weyl fermions in candidate topological semimetals.
\end{abstract}

\maketitle

Until groundbreaking experiments regarding the two-dimensional material graphene, a study of relativistic quasiparticles has been limited to the high-energy physics  \cite{Novoselov}. This single-layer allotrope of carbon is a zero-gap semiconductor with a linear energy dispersion of conduction and valence bands connected one with the other at their extremities, and thus giving rise to the presence of low-energy quasiparticles governed by the relativistic Dirac equation \cite{Neto}. Even more promising for quantum information processing are certain three-dimensional semimetals with non-trivial topology that host  massless chiral fermions as quasiparticle excitations described by the relativistic Weyl equation \cite{Xu,Yan}. Due to the breaking of either the inversion symmetry or the time reversal symmetry, a Weyl semimetal is characterized by the band-touching points known as Weyl nodes around which the singly degenerate bands disperse linearly in all three momentum-space directions  \cite{Armitage}.

Because of the inherent chirality of Weyl quasiparticles and the emerged monopole-like structure of the Berry curvature, Weyl semimetals promise the wealth of novel phenomena. In particular,  the surface Fermi arcs, that are directly observed using momentum-resolved photoemission spectroscopy, are recognized as a prime characteristic of this class of topological semimetals \cite{Armitage}. The negative longitudinal magnetoresistance \cite{niel,son,para,zhang-nonl} caused by the chiral anomaly, and unusual quantum oscillations produced by a cyclotron motion that weaves together the Fermi arcs and chiral bulk states \cite{pott,moll} are  other inherent properties of these materials.  Recently, a unique type of acoustic collective mode called chiral zero sound has been theoretically proposed for Weyl semimetals with multiple pairs of Weyl nodes \cite{Song}, and giant quantum oscillations of the thermal conductivity discovered in the prototypical Weyl semimetal TaAs have been explained with this chiral sound \cite{Xiang}. However, most of these experimental signatures of the Weyl electrons are often difficult to track down, in particular because the relativistic fermions coexist with conventional quasiparticles in  topological semimetals. For example, although the negative longitudinal magnetoresistance in a parallel magnetic field was observed in a number of Weyl semimetals, its interpretation as the long-sought manifestation of the chiral anomaly remains controversial due to a possible inhomogeneous current flow in bulk crystals  \cite{Reis,Ramshaw} as well as in view of alternative explanations of this effect \cite{Naumann}.  At present, measurements of a quantum-oscillation phase are widely used to detect the Weyl fermions (see, e.g., references in review \cite{m-sh19}) since this phase is noticeably affected  by the Berry curvature.
However, such experiments sometimes lead to ambiguous results.
Therefore, and because of a rapidly growing number of candidate materials, new experimental methods to detect signatures of relativistic quasiparticles in topological semimetals are highly desirable. In this context, it was recently shown that the magnetization and magnetic torque measurements of Weyl semimetals upon entering the ultra-quantum limit state in high magnetic fields
can be a useful probe for discerning the relativistic quasiparticles
\cite{Moll,Zhang,modic}.

In this article, we draw  attention to the magnetostriction, i.e., to the field-induced length change, which results from the interaction between the electron and elastic degrees of freedom in a crystal.
Using TaAs as an example, we show that measuring this thermodynamic quantity, one can clearly distinguish between the relativistic and conventional electrons already in the field range where the Weyl fermions are confined at their zeroth Landau level, but the trivial quasiparticles are far below their ultra-quantum limit. Developing a theory of the magnetostriction for topological semimetals with a noncetrosymmetric crystal structure, we demonstrate how parameters characterizing not only Weyl electrons but also their interaction with elastic deformations can be extracted from the magnetostriction measurements. A firm evidence for Weyl fermions is found along the [001] direction where the largest length changes are observed. By contrast, the longitudinal expansion along the [100] direction is by an order of magnitude smaller in the highest field applied, but this $a$-axis magnetostriction experiences immense changes from large positive to large negative values with minute deviations of the applied magnetic field from the [001] direction. We suppose that the observed anisotropic magnetostrictive stress can be relevant for future high-field Weyltronic devices.

\section{Results}

\textbf{Magnetostriction of nonmagnetic semimetals.}
The magnetostriction of nonmagnetic conductive materials is directly related to changes in the density of charge carriers in a magnetic field. Specifically, a pocket $i$ of the Fermi surface makes the following contribution to the field-induced relative length change (Supplementary Note $1$):
\begin{eqnarray}\label{1}
\frac{\Delta L}{L}\,=\,\Lambda_i \left(n_i(B)-n_i(0)\right),
\end{eqnarray}
where $B$\,=\,$\mu_0H$ is the magnetic induction produced in the sample by the external magnetic field $H$, $n_i(B)$ is the $B$\nobreakdash-dependent density of the charge carriers in  this pocket, and the constant $\Lambda_i$ depends on the direction along which the magnetostriction is measured. Formula (\ref{1}) results from a minimization of the energy consisting of the elastic energy proportional to $(\Delta L/L)^2$ and of the energy of the interaction between the elastic and electron degrees of freedom. This formula is written under the assumption that a deformation of the crystal shifts the appropriate electron band as a whole and does not change its shape. As a rule, this rigid-band approximation is quite accurate for real semimetals. Indeed, it was experimentally shown that the magnetostriction is very small if all charge carriers belong to a single band \cite{Mi}, and this small value characterizes the precision of the rigid-band approximation. [In this case, formula (\ref{1}) predicts that the magnetostriction vanishes since $n_i(B)$\,=$\,n_i(0)$ due to the conservation of the carriers]. On the other hand, for a multiband material, this thermodynamic quantity  is greatly enhanced \cite{Mi}  due to a band overlap and an electron redistribution between the bands at the switching-on of the magnetic field. Below we consider only such multiband materials since all the known Weyl semimetals contain several groups of the charge carriers.

\begin{figure}[tbp] 
 \centering  \vspace{+9 pt}
\includegraphics[scale=0.54]{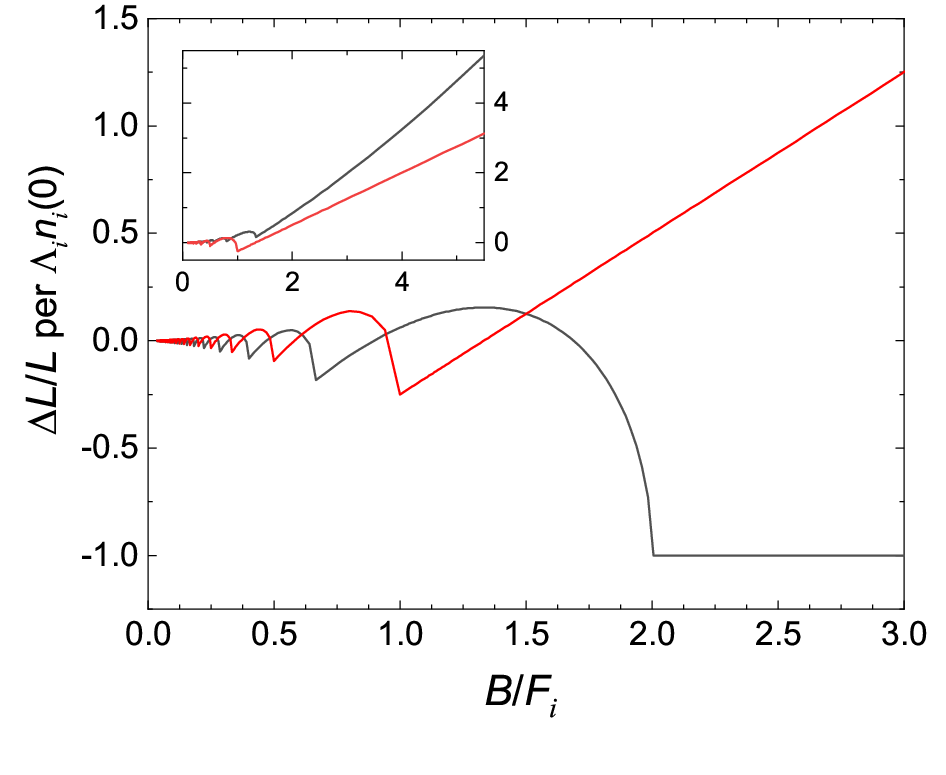}
\caption{\label{fig1} \textbf{The magnetostriction of the Weyl electrons (red) and of the trivial electrons with parabolic spectrum (black) versus magnetic induction $B=\mu_0H$ at zero temperature}. Here $F_{i}$ is defined by Eq.~(\ref{3}), the magnetostriction $\Delta L/L$ is calculated with equations of Supplementary Notes 2 and 3. For the trivial electrons, the parameter $\delta$ is assumed to be equal to zero. Note the relative phase shift of the oscillations shown by the red and black curves. Inset: The similar plot, but when $\delta$\,=\,0.75 for the trivial electrons (black line). In this case $\Delta L/L=\Lambda_i n_i(0)(0.75u^{-3/2} \sqrt{u+\delta-0.5}-1)$ for $B$\,$>$\,$F_{i}/(1.5-\delta)$ where $u\equiv F_{i}/B$ (Supplementary Note 3).
 } \end{figure}

\textbf{Distinctions between the Weyl and trivial electrons.}
When several groups of electrons or holes exist in a conductive material, their Fermi energy $E_F$ (or the chemical potential $\zeta$ at nonzero temperature) generally depends on the magnetic field. At first, however, we will neglect this $B$ dependence of $E_F$ since as will be shown below, this simplified approach can provide a sufficiently accurate description of the magnetostriction. We start with a comparison of the magnetostrictions produced by the Weyl  quasiparticles and by the trivial electrons for which the spectrum has the parabolic form (Supplementary Notes $2$ and $3$). For trivial electrons in high magnetic fields, their lowest Landau level  rises above the Fermi energy if the parameter $\delta$ characterizing the electron magnetic moment $\mu_e$\,=\,$\delta (e\hbar/m_*)$ is less than $1/2$ where $m_*$ is the cyclotron mass. This moment consists of its spin and orbital parts, the latter being due to the spin-orbit interaction. For such fields, the Fermi-surface pocket $i$ of the  trivial electrons empties, and $n_i(B)$\,=\,0 in this ultra-quantum limit. Thus, the magnetostriction of these electrons becomes constant,
\begin{eqnarray}\label{2}
\frac{\Delta L}{L}=-\Lambda_i n_i(0)\equiv a_i.
\end{eqnarray}
This saturation of the magnetostriction takes place at $B$\,$>$\,$F_{i}/(0.5-\delta)$, see Fig.~\ref{fig1}, where
\begin{equation}\label{3}
F_{i}=\frac{S_{{\rm max},i}}{2\pi e\hbar}
\end{equation}
is frequency of the quantum oscillations occurring at $B< F_{i}$, and $S_{{\rm max},i}$ is the maximal cross-sectional area of the pocket $i$. On the other hand, if $\delta>1/2$, the orbital $B$-dependent  displacement $(e\hbar/2m_*)B$ of the lowest Landau level from the band edge $\varepsilon_0$ is less than the electron Zeeman energy $\mu_eB$, and at least one  Landau level remains occupied by the charge carriers at any $B$. In this case,
\begin{equation}\label{4}
\frac{\Delta L}{L}\approx a_i+\gamma_i  B^{3/2},
\end{equation}
for the magnetic fields $H\gg F_{i}/\mu_0$, with the factor $\gamma_i$ being dependent on $\delta$ (Fig.~\ref{fig1} and Supplementary Note 3).

In the weak magnetic fields $H\ll F_{i}/\mu_0$, if the quantum oscillations are suppressed by impurities or a temperature, the magnetostriction takes the form
\begin{eqnarray}\label{5}
\frac{\Delta L}{L}=-a_i\frac{3 B^2}{8F_{i}^2}\left(\delta^2-\frac{1}{12}\right)\equiv c_iB^2.
\end{eqnarray}
If the oscillations are superimposed on a smooth background, this formula just describes this background.

In the case of the Weyl electrons, the zero Landau level coincides with the energy of the Weyl point for any $B$. Therefore, in the ultra-quantum regime when $\mu_0H$\,$>$\,$F_{i}$, only this level is occupied by the electrons, their density $n_i(B)$ is proportional to $B$, and the magnetostriction is described by the formula
\begin{eqnarray}\label{6}
\frac{\Delta L}{L}=a_i\left(1- \frac{3B}{4F_{i}}\right) \equiv a_i+b_iB,
\end{eqnarray}
where $a_i$\,=\,-$\Lambda_i n_i(0)$, the parameter $F_{i}$ is still defined by formula (\ref{3}) and coincides with the frequency of the quantum oscillations. A comparison of Eq.~(\ref{6}) with formulas (\ref{2}) and (\ref{4}) for the parabolic spectrum shows that in the high-field region, the magnetostriction produced by Weyl fermions  essentially differs from the magnetostriction of the trivial electrons (see also Supplementary Note 3).

For $H$\,$\ll$\,$F_{i}/\mu_0$,  the magnetostriction of the Weyl  electrons takes the form coinciding with Eq.~(\ref{5}) at the demarcative value of $\delta$\,=\,1/2,
\begin{eqnarray}\label{7}
\frac{\Delta L}{L}=-\frac{a_i B^2}{16F_{i}^2}\equiv c_iB^2.
\end{eqnarray}

It is clear from formulas (\ref{2})--(\ref{7}) that fits of the quadratic function $c_bB^2$ to the smooth background in the weak-field range  and of a linear polynomial or a power function  to the magnetostriction in the ultra-quantum regime allow one not only to detect the Weyl electrons but also to determine at least a part of the parameters $a_i$ and $F_{i}$ from the experimental data. With these parameters, the magnetostriction, including its subtle details as the quantum oscillations, can then be calculated in the entire range of the applied magnetic fields, using  formulas of Supplementary Notes 1-3. A comparison of results of this calculation with the appropriate experimental data permits one to verify the  existence of the conjectured Weyl electrons and to refine the values of the parameters for them.

\textbf{Advantages of the magnetostriction.}
Consider a dependence of the magnetostriction on the volume of the Fermi-surface pockets (Supplementary Notes $2$ and $3$). In the case of the trivial electrons, one has
$n_i(0)\propto |E_F-\varepsilon_0|^{3/2}$, $F_{i}\propto S_{{\rm max},i}\propto |E_F-\varepsilon_0|$, and Eq.~(\ref{5}) yields for the weak magnetic fields,
\begin{eqnarray}\label{8}
c_i&\propto &|E_F-\varepsilon_0|^{-1/2}\propto n_i^{-1/3},\ \ \ \ \ \ E_F>\varepsilon_0, \\
c_i&=&0,\ \ \ \ \ \ \ \ \ \ \ \ \ \ \ \ \ \ \ \ \ \ \ \ \ \ \ \ \ \ \ \ \ \ \ \ \  E_F<\varepsilon_0, \nonumber
\end{eqnarray}
where $\varepsilon_0$ is the edge of the energy band. (For the trivial holes the same formulas hold true but at the opposite relations between $E_F$ and $\varepsilon_0$). A similar increase of the coefficient $c_i$ with decreasing the charge-carrier density $n_i$ is obtained for the Weyl electrons from Eq.~(\ref{7}) since $n_i(0)\propto (E_F-\varepsilon_d)^{3}$, $F_i\propto S_{{\rm max},i}\propto (E_F-\varepsilon_d)^2$ in this case, and hence
 \begin{eqnarray}\label{9}
c_i&\propto &\pm |E_F-\varepsilon_d|^{-1}\propto n_i^{-1/3},
 \end{eqnarray}
where  $\varepsilon_d$ is the energy of the Weyl point, and the signs $\pm$ correspond to the electrons ($E_F$\,$>$\,$\varepsilon_d$) and holes ($E_F$\,$<$\,$\varepsilon_d$), respectively. The obvious difference between the Weyl and trivial charge carriers is that the function  $c_i(E_F)$ changes its sign at the energy of the Weyl point $\varepsilon_d$, whereas this change does not occur for the trivial fermions. More importantly, however, the relation $c_i$\,$\propto$\,$n_i^{-1/3}$ reveals that the magnetostriction is substantially larger for small electron pockets than for large electron groups. That is why this quantity for elemental bismuth  ($\Delta L/L\gtrsim 10^{-6}$ at 10 T \cite{Mi,Kuchler1}), the Fermi surface of which consists of small pockets, considerably exceeds the magnetostriction of metals ($\Delta L/L\sim 10^{-8}$ at 10 T \cite{fawcett}). This feature of the magnetostriction simply reflects the fact that the change of charge-carrier density in the weak magnetic fields (and at constant $E_F$) is less for a large Fermi-surface pocket than for a small one. In the ultra-quantum regime, when the change in the density becomes of the order of the density itself, the contribution to the  magnetostriction generated by the large electron group can eventually exceeds the appropriate contribution of the small one, but such extreme fields are not currently available for dilatometric experiments.

Compare now the magnetostriction with those physical quantities that are proportional to the density of charge-carrier states at $B=0$, $\nu_i$\,=\,$dn_i(E_F)/dE_F$ (e.g., with the non-oscillating part of the electrical conductivity \cite{abr}). This $\nu_i$ increases with $n_i$ both for the trivial electrons ($\nu_i\propto n_i^{1/3}$) and for the Weyl quasiparticles ($\nu_i\propto n_i^{2/3}$). Hence, in measurements of those quantities,  experimental signatures for small Fermi-surface pockets are extensively masked by the contribution of a large pocket when it exists in the material. Therefore, the above $n_i$ dependence of the magnetostriction makes it a useful tool for studying topological semimetals in which the Weyl points are in the vicinity of $E_F$.

In Supplementary Note 4 we also compare the magnetostriction with the magnetization $M$, the  orbital part of which is not determined by the density of electron states as well, and which is considered as another thermodynamic probe of the Weyl electrons \cite{Moll,Zhang,modic}. This comparison reveals the following distinctions between these quantities: i) Completely filled electron bands and, as clear from the above considerations, large Fermi pockets practically do not contribute to the magnetostriction. However, such pockets and even filled energy bands produce most of the magnetization, and this part of $M$ remains  proportional to $B$ for the magnetic fields at which the Weyl electrons are in the ultra-quantum regime. Therefore, it is necessary to carry out a subtraction of extrapolated low-field magnetization from the high-field experimental data in order to extract the Weyl-electron contribution to $M$ \cite{m-sh16,m-sh19}. ii) If only small charge-carrier pockets exist in a semimetal, all these pockets can make large contributions to the magnetostriction. On the other hand, an attractive feature of the magnetization is that only its part produced by Weyl points is relatively large, whereas the part generated by small trivial-electron groups is insignificant.
iii) Although the magnetostriction and the magnetization are similar in many  respects, these quantities are associated with the different parts of the free energy of the conductive materials. The magnetization characterizes the electron energy in a magnetic field, whereas the magnetostriction results from the sum of the elastic energy and the energy of the interaction between the electron and elastic degrees of freedom in a crystal. For this reason, the detailed analysis of the magnetization and the magnetostriction can provide  complementary information on the parameters of the Weyl points. In particular, the magnetostriction depends not only on the electron characteristics $F_i$, $n_i(0)$, but also on the constants $\Lambda_i$ which are determined by the elastic moduli of the crystal and by the constants of the deformation potential (Supplementary Note 1). These constants specify shifts of the energy bands under  deformations in the conductive materials.

A measurement of the magnetic torque is the effective way of determining the transverse component of the magnetization \cite{Moll,Zhang,modic}. However, this component and the magnetic torque  vanish when the magnetic fields is aligned with a symmetry axis of a crystal. At magnetic fields tilted away from the symmetry axis, even equivalent pockets in a Weyl semimetal  produce different contributions to the magnetization and to the magnetostriction, and a theoretical analysis of these quantities becomes complicated for the semimetals with multiple Weyl nodes. However, the magnetostriction as well as the longitudinal magnetization and the magnetotropic coefficient \cite{modic18} remain nonzero at the magnetic field aligned with the symmetry axis, and such measurements of these quantities seem to be most convenient for an initial analysis of the Weyl fermions.

\textbf{Magnetostriction of TaAs along the [001] direction.}
Let us exemplify the above considerations by an investigation of the field-induced length change of the Weyl semimetal TaAs.
Like other transition-metal monopnictides NbP, NbAs, and  TaP \cite{klotz,komada,naumann1,arnold1}, tantalum arsenide crystallizes in a body-centered tetragonal structure that lacks a horizontal mirror plane and thus the inversion symmetry. This noncentrosymmetric structure is essential to the existence of multiple pairs of the Weyl nodes divided into four pairs of the W1 points and eight pairs of the W2 points \cite{Xu}. Among these materials, TaAs exhibits the largest separation of the Weyl nodes in momentum space, and the Fermi energy is sufficiently close to the Weyl points to produce a separate Fermi pocket encompassing each of these points. In addition to this, the cross-sectional areas of both the W1 (banana shaped) and W2 (nearly isotropic sphere) pockets are so small that the ultra-quantum limit for the Weyl electrons can be easily reached in experiments  \cite{Arnold,Ramshaw}. These electron pockets coexist with trivial hole pockets aligned along the nodal rings which would occur in TaAs if the spin-orbit interaction were absent \cite{Arnold}.
When the magnetic field is parallel to the $c$ axis of TaAs, all the pockets in each of the W1 and W2 electron groups or in the group of the holes have equal densities $n_i(B)$ and extremal cross-sectional areas $S_{{\rm max},i}$. Let  $F_{W1}$, $F_{W2}$, and $F_{h}$ denote the frequencies of quantum oscillations produced by the W1 and W2 electrons and by the holes, respectively. As discussed above, these $F_i$ also correspond to the magnetic inductions at which the W1 and W2 electrons and the holes enter the ultra-quantum regime. According to Ref.~\cite{Arnold}, one has $F_{W1}$\,$\simeq$\,7\,T, $F_{W2}$\,$\simeq$\,5\,T, and $F_{h}$\,$\simeq$\,19\,T, with $F_{W2}$ being calculated but not measured for the [001] direction of the magnetic field. We emphasize that for $B$\,$\parallel$\,$c$, the entire field range up to $16$ T, which is available in our experiments, may be considered as the low-field region for the holes.

\begin{figure}
	\begin{center}
\includegraphics[scale=0.54]{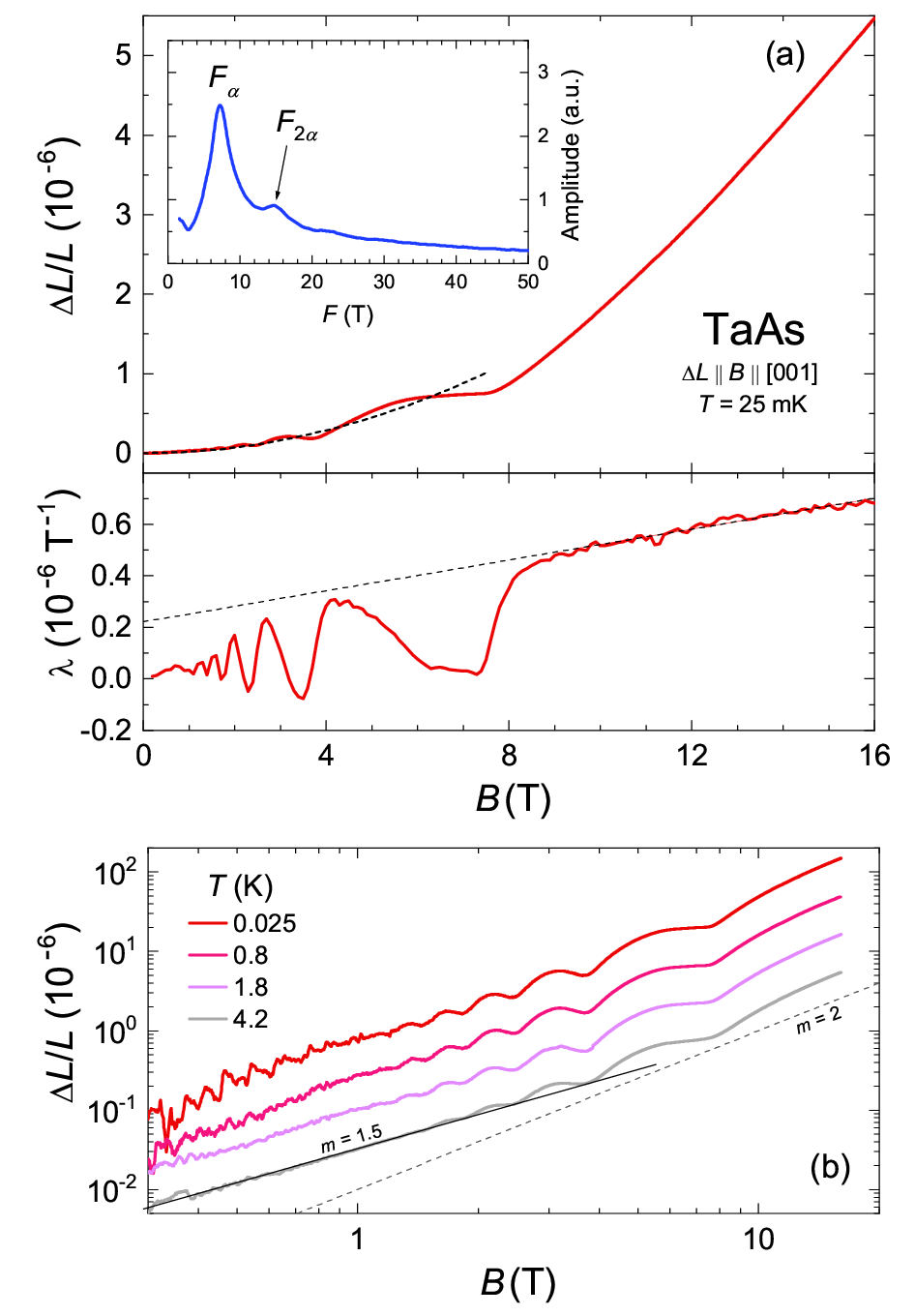}
		\caption{\textbf{Magnetostriction in the Weyl semimetal TaAs.} (a) Top: Magnetic-field dependence of the relative length change {$\Delta L/L$} of TaAs (sample\,1) measured along the [001] direction at 25\,mK in the parallel configuration ($B$\,$\parallel$\,$\Delta L$). Inset:
FFT spectrum at $T$\,=\,25\,mK over a magnetic field range of 1 to 8\,T. The oscillatory magnetostriction was obtained by subtracting the background $c_bB^2$ with $c_b$\,=\,1.79\,$\times$\,10$^{-8}$\,T$^{-2}$ (dashed line in the main panel) from the experimental data.
Bottom: Corresponding magnetostriction coefficient $\lambda$\,$\equiv$\,$(1/L)d L/dB$.
The straight dashed line approximating $\lambda(B)$ gives the intercept ($2.28\times 10^{-7}$ T$^{-1}$) and has the slope ($2.95\times 10^{-8}$ T$^{-2}$) which are close to the parameters $b$ and $2c_h$ in the quadratic polynomial $a+bB+c_hB^2$ (see the text and Table I). (b) A log-log graph of the $c$-axis magnetostriction measured at different temperatures up to 4.2\,K. Curves are offset for clarity. Note the slopes clearly distinct from the quadratic ($m$\,=\,2) behavior expected for  $B$\,$\ll$\,$F_i$. The dashed and solid black  straight lines correspond to $\Delta L/L\propto B^{m}$.}
		\label{fig2}
	\end{center}
\end{figure}

Figure \ref{fig2}(a) shows the magnetostriction of TaAs measured along the [001] direction at 25\,mK. The field-induced expansion  is large and the relative length change $\Delta L/L$ amounts to about 5.5\,$\times$\,10$^{-6}$ at $B$\,=\,16\,T. With the magnetic field aligned along the $c$ axis, the quantum oscillations reaching large  amplitudes  ($\sim$\,30\,\% of the background signal at\,3\,T) are observable in the raw {$\Delta L/L$}  data (top panel) and even more sharply discernible in the derived coefficient $\lambda$\,=\,$\frac{1}{L}\frac{dL}{dB}$ (bottom panel) that represents the derivative of the charge-carrier density with respect to $B$. The sharp change observed in $\lambda$ at $\sim$7.5\,T and the  linearly increasing signal at higher fields clearly show that the W1 and W2 electron groups enter the ultra-quantum regime, and  the magnetostriction above 8\,T is well approximated by the square polynomial $a+bB+c_hB^2$. The fit gives the values of $a$, $b$, and $c_h$ presented in Table I. The presence of the linear-in-$B$ term in the magnetostriction clearly indicates the existence of the Weyl electrons, whereas the term $c_hB^2$ may be associated with the holes that are in the low-field range at $B\le 16$\, T. Thus, the presented experimental data do  demonstrate the possibility of detecting the  relativistic fermions with the magnetostriction.

Finally, we note that except for the oscillatory features, the overall $B$-dependence of the  $c$-axis magnetostriction remains  essentially unaltered up to the temperature 4.2\,K [cf. Fig.~\ref{fig2}(b)]. However, the log-log graph seems to reveal a deviation  from the $B^2$ law expected in the low-field region (dashed black line). This finding might point to  a very small value of $F_{W2}$ for $B\parallel$\,[001]. However, the deviation of the $B$-dependence from the quadratic behavior is less than or of the order of $10^{-8}$, and hence a detailed insight into the low-field region requires larger TaAs crystals.

\begin{table}\label{tab1}
\caption{\textbf{The values of the parameters for the calculation of the magnetostriction $\Delta L/L$ along the $c$ axis at $B\parallel c$ in neglect of the $B$ dependence of $\zeta$.}  Values of $a_{W1}$ and $a_{W2}$ are found with Eqs.~(\ref{10}) from the coefficients of the polynomial $a+bB+c_hB^2$ that approximates the experimental data on the magnetostriction at $B>8$ T.}
\begin{tabular}{cccc|cccc|cc}
\hline
\hline \\[-2.5mm]
$a$&$b$&$c_h$&$F_{W1}$&set&$F_{W2}$&$a_{W1}$&$a_{W2}$&$\gamma_{W1}$&$\gamma_{W2}$ \\
$10^{-6}$&$10^{-7}$T$^{-1}$&$10^{-8}$T$^{-2}$&T&\#&T&$10^{-6}$&
$10^{-6}$&~&~ \\
\colrule
$-2.18$&$2.55$&$1.49$&$7.2$&$1$&$5$&$-1.58$&$-0.60$&$0.025$&$0.1$ \\
~&~&~&~&$2$&$1.35$&$-2.12$&$-0.062$&~&~ \\
\hline \hline
\end{tabular}
\end{table}

\begin{figure}[tbp]
 \centering
\includegraphics[scale=0.54]{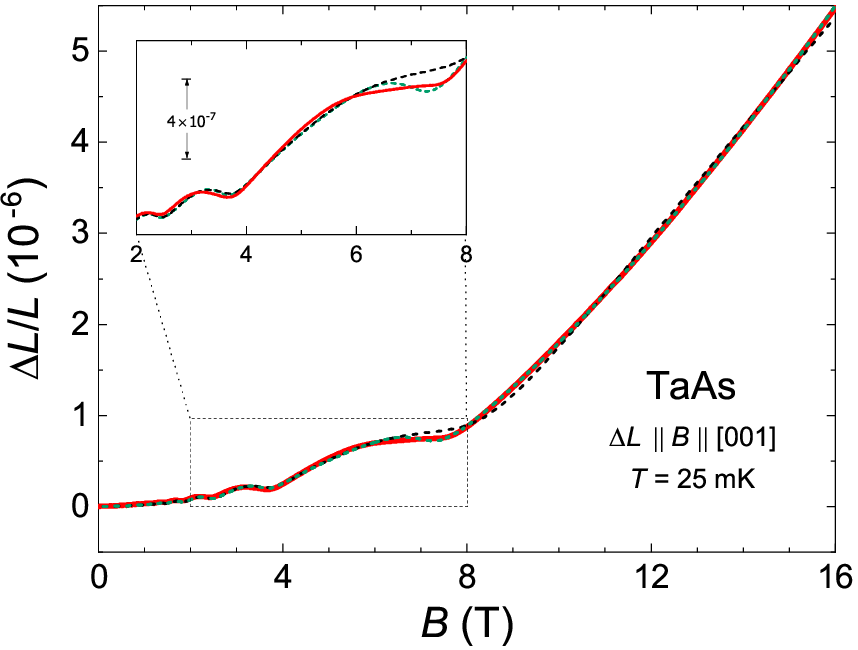}
\caption{\label{fig3} \textbf{Comparison of the calculated magnetostriction for the case of $F_{W2}=5$ T with the experimental data  for TaAs.}
The solid red line shows the $c$-axis magnetostriction $\Delta L/L$ of  TaAs measured at $T$\,=\,25\,mK and $B\parallel c$.
The dashed green line depicts this magnetostriction calculated at $T=0$ with formulas of Supplementary Notes $1-3$ (assuming the constancy of the chemical potential $\zeta$) for the values of the parameters presented in Table I (set 1). The dashed black line shows the magnetostriction calculated in Supplementary Note $7$ for $T=0$ and for the same values of $F_i$ and $\gamma_i$, taking into account the dependence $\zeta(B)$. The inset is a zoom into the last oscillations of the magnetostriction. Note the absence of a local minimum in the experimental curve above 6 T.
 } \end{figure}

\textbf{Analysis of the magnetostriction for TaAs.}
The theory of the magnetostriction presented in Supplementary Notes $1-3$, permits us to describe quantitatively the experimental data for TaAs. We find that at $F_{W1}$\,=\,7.2\,T, the frequency of the calculated oscillations in the magnetostriction coincides with that observed experimentally. Since this frequency agrees with  $F_{W1}$\,=\,7\,$\pm$\,0.5\,T reported by Arnold et al. \cite{Arnold}, we may be guided by that work in our analysis. As in all previous experiments with TaAs \cite{Xiang,Arnold,Zhang,laliberte}, the oscillations with the frequency $F_{W2}$ do not manifest themselves in our data on the magnetostriction, and we take $F_{W2}$\,=\,5\,T to maintain agreement with the results of the band-structure calculations \cite{Arnold}.  Then, using formula (\ref{6}) and the values of the constants $a$ and $b$ found above from the approximation of $\Delta L/L$ at $B$\,$>$\,$8$\,T, we arrive at the two linear equations in the parameters $a_{W1}$ and $a_{W2}$ characterizing the W1 and W2 electrons,
 \begin{eqnarray}\label{10}
 a_{W1}+a_{W2}=a, \\
 -0.75\left(\frac{a_{W1}}{F_{W1}}+\frac{a_{W2}}{F_{W2}} \right)=b. \nonumber
 \end{eqnarray}
These equations give $a_{W1}\approx-1.58\times 10^{-6}$, $a_{W2}\approx -0.60\times 10^{-6}$ (set 1 in Table I).
Note that at given $F_{W1}$ and $F_{W2}$, we have been able to determine all the unknown parameters for TaAs since there are only two nonequivalent  groups of the Weyl electrons in this semimetal. Interestingly, the obtained $a_{W1}$ and $a_{W2}$ predict the value of the coefficient $c_b$ determining the low-field behavior $c_bB^2$ of the magnetostriction,
\begin{eqnarray*}
c_b=c_h -\frac{a_{W1}}{16F_{W1}^2}-\frac{a_{W2}}{16F_{W2}^2} \approx 1.83\times 10^{-8}\,{\rm T}^{-2},
   \end{eqnarray*}
and this coefficient is close to that found experimentally (cf. the dashed line in Fig.~\ref{fig2}(a), top). With these parameters and with  formulas of Supplementary Notes $1-3$ at zero temperature, we calculate the magnetostriction of the Weyl electrons for all $B$\,$\le$\,16\,T. To fit the  magnitude of calculated oscillations to the experimental data, we use the dimensionless parameter specifying the scattering of the W1 electrons by impurities,  $\gamma_{W1}$\,=\,$\pi T_{D,W1}/(E_F-\varepsilon_{d,W1})$, and find from the fit that $\gamma_{W1}$\,=\,0.025  where $\varepsilon_{d,W1}$ is the energy of the Weyl points W1, and $T_{D,W1}$ is  the Dingle temperatures for the W1 electrons. The similar ratio $\gamma_{W2}=\pi T_{D,W2}/(E_F-\varepsilon_{d,W2})$ for the W2 electrons is assumed to be equal to $0.1$ in order to suppress the appropriate oscillations. Adding the hole contribution $c_hB^2$ to the calculated magnetostriction of the Weyl electrons, we find that the theoretical curve sufficiently well reproduces the experimental data in entire field range up to  $16$ T (compare the dashed green and solid red lines in Fig.~\ref{fig3}) except for the $6-7.5$ T interval where the last oscillation sets in (see the zoom in Fig.~\ref{fig3}).

The above theoretical analysis  of the magnetostriction is based on formulas obtained under the assumption of independence of the Fermi energy on the magnetic induction. This situation does can occur in a conductive material when it contains a large  charge-carrier group that maintains the constancy of $E_F$. However, in TaAs all the electron and hole pockets are relatively small. In this case, a consideration must be given to the $B$ dependence of $E_F$ (i.e., of the chemical potential $\zeta$ if the temperature is nonzero) in analyzing the magnetostriction. This dependence $\zeta(B)$ can be found from the  conservation condition of the total  charge-carrier density,
\begin{eqnarray}\label{11}
\sum_i \left(n_i(\zeta,B)-n_i(\zeta_0,0)\right)=0,
\end{eqnarray}
where $i$ runs all the electron and hole pockets, and $\zeta_0$ is the value of the chemical potential at $B$\,=\,0.
Since the dispersion law for the holes need not be well described by a simple parabolic dependence, we use the expression for $n_h(\zeta,B)- n_h(\zeta_0,0)$ that is valid at $B$\,$<$\,$F_{h}$ for any dispersion of these charge carriers,
\begin{eqnarray*}
 n_h(\zeta,B)\!&-&\!n_h(\zeta_0,0)\!=\!n_h(\zeta,B)\!-\!n_h(\zeta,0)
 +n_h(\zeta,0)\!-\!n_h(\zeta_0,0)\nonumber \\
&\approx&
B^2\Big(\beta(\zeta_0)+\frac{d\beta(\zeta_0)}{d\zeta_0} (\zeta-\zeta_0)\Big)+\nu_h(\zeta_0)(\zeta-\zeta_0),~~~
 \end{eqnarray*}
where  $\nu_h$\,=\,$\partial n_h(\zeta,0)/\partial \zeta$ is the density of states for the holes in zero magnetic field, whereas the function $\beta(\zeta)$ defines the variation of the hole density in the low  magnetic fields, $n_h(\zeta,B)- n_h(\zeta,0)$\,=\,$\beta(\zeta)B^2$ [i.e.,  $\beta$\,=\,$c_h/\Lambda_h$ where $c_h$ has been introduced above].

\begin{figure}[tbp] 
 \centering  \vspace{+9 pt}
\includegraphics[scale=0.54]{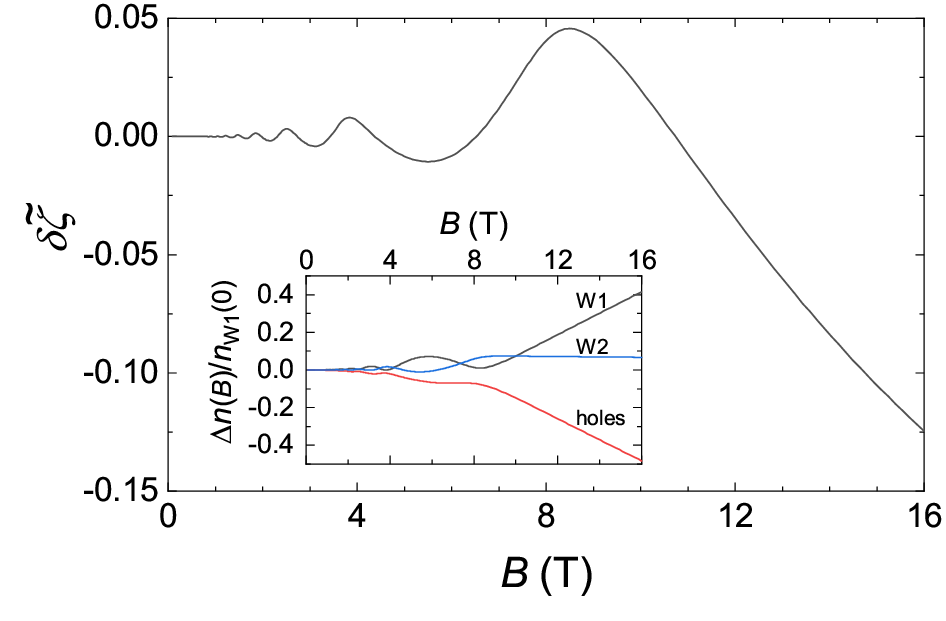}
\caption{\label{fig4} \textbf{The field-induced shift of the chemical potential  relative to its value $\zeta_0\equiv \zeta(0)$ at zero magnetic field in  TaAs}. Shown is the normalized shift  $\delta\tilde\zeta\equiv (\zeta(B)-\zeta_0)/(\zeta_0-\varepsilon_{W1})$ that is calculated together with the dashed black line in Fig.~\ref{fig3}. Inset: Dependences of $(n_{W1}(B)-n_{W1}(0))/n_{W1}(0)$, $(n_{W2}(B)-n_{W2}(0))/n_{W1}(0)$, and $(n_{h}(B)-n_{h}(0))/n_{W1}(0)$ on $B$. Note that although $\delta\tilde\zeta$ decreases with increasing field above $8.5$\,T, $n_{w1}(B)$ simultaneously increases due to the growing capacity of the Landau levels.}
\end{figure}

Equation (\ref{11}) and the general formula for the magnetostriction are explicitly written in Supplementary Note $5$  for the case of  $B$\,$\parallel$\,$c$, and with these expressions, one can calculate the magnetostriction in the entire range of the applied magnetic fields. As in the case of the simplified approach when $\zeta(B)$\,=\,$\zeta_0$, we set $F_{W1}$\,=\,7.2\,T, $F_{W2}$\,=\,5\,T, and take the same values of the constants $\gamma_{Wi}$. Apart from these parameters, the chemical potential $\zeta(B)$  depends also on the ratios $n_{W2}(\zeta_0,0)/n_{W1}(\zeta_0,0)$, $(\zeta_0-\varepsilon_{d,W1})/(\zeta_0-\varepsilon_{d,W2})\equiv v$, and on the above-mentioned  $\beta$, $d\beta/d\zeta$, $\nu_h$ normalized to $n_{W1}(\zeta_0,0)$ where $\varepsilon_{d,W1}$ and $\varepsilon_{d,W2}$ are the energies of the Weyl points W1 and W2. Applying formulas of Ref.~\cite{m-sh21a} to the data of Ref.~\cite{Arnold}, we may estimate a part of these parameters, viz., the density $n_{W1}$\,$\approx$\,2.5\,$\times$\,10$^{18}$\,cm\,$^{-3}$, the ratio $n_{W2}(\zeta_0,0) /n_{W1}(\zeta_0,0)$\,$\sim$\,0.15,  and the position of the chemical potential $\zeta_0$ relative to the energies $\varepsilon_{W1}$, $\varepsilon_{W2}$: $\zeta_0-\varepsilon_{d,W1}$\,$\approx$\,28.4\,$\pm\,3.5$\,meV and  $\zeta_0-\varepsilon_{d,W2}$\,$\approx$\,11.9\,$\pm$\,1\,meV  (Supplementary Note $6$). These values of the parameters permit us to set $n_{W2}(\zeta_0,0) /n_{W1}(\zeta_0,0)$\,=\,0.15 and $v$\,=\,2.5  in our calculations of the magnetostriction.
At these fixed ratios, the values of the other parameters are chosen so that the  magnetostriction calculated at $T=0$ matches the experimental data at $T=25$ mK (Fig.~\ref{fig3} and Supplementary Note $7$). Note that with the dependence $\zeta(B)$, the theoretical curve much better reproduces the plateau above about $6$ T than in the case of the constant $\zeta$. The derived dependence of the chemical potential on the applied magnetic field is presented in Fig.~\ref{fig4}. It is seen that due to condition (\ref{11}), the largest electron group W1 induces the oscillation with the same frequency $7.2$ T for the other charge carriers. The analysis of the obtained parameters in Supplementary Note 7 enables us to find the constants $\Lambda_{i}\equiv \Lambda_{i}^c$, which determine contributions of the W1 and W2 electrons and of the holes to the $c$-axis magnetostriction of TaAs (Table II). This analysis also reveals that the obtained values of $\beta$, $d\beta/d\zeta$, and $\nu_h$  can be understood from simple estimates, assuming the simplest parabolic  dispersion of the holes in TaAs. We also obtain a good fit of the magnetostriction calculated at a finite dimensionless temperature $t$\,=\,$T/(\zeta_0-\varepsilon_{W1})=0.015$  to the magnetostriction measured at the temperature $T$\,=\,4.2\,K (Supplementary Fig.~8) and therefore find the independent  estimate of  $\zeta_0-\varepsilon_{W1}$\,$\approx$ \,24\,meV, which is only a little less than the value 28.4\,$\pm$\,3.5\,meV derived above. With this $\zeta_0-\varepsilon_{W1}$ and $\gamma_{W1}=0.025$, we arrive at the Dingle temperature $T_{D,W1}=(\zeta_0- \varepsilon_{W1}) \gamma_{W1}/\pi\approx 2.2$\,K, the value of which
is comparable with $T_{D,W1}$\,$\approx$\,3.2\,K obtained for the W1 electrons in Ref.~\cite{Arnold}. (As to $T_{D,W2}$, we tentatively find $T_{D,W2}$\,$\sim$\, 3.6\,K; Supplementary Note $7$.)

\begin{table}
\caption{\textbf{The values of $\Lambda_{i}^c$ and $\Lambda_{i}^{\perp}$ obtained from the $c$-axis and $a$-axis magnetostrictions, respectively.}}
\begin{tabular}{cc|cccccc}
\hline
\hline \\[-2.5mm]
set&$F_{W2}$&$10^{24}\Lambda_{W1}^c$&$10^{24}\Lambda_{W2}^c$&
$10^{24}\Lambda_{h}^c$&
$10^{24}\Lambda_{W1}^{\perp}$&$10^{24}\Lambda_{W2}^{\perp}$&
$10^{24}\Lambda_{h}^{\perp}$ \\
 \#&T&cm$^{3}$&cm$^{3}$&cm$^{3}$&
 cm$^{3}$&cm$^{3}$&cm$^{3}$  \\
\colrule
$1$&$5$&$0.71$&$1.47$&$-3.7$&~&~&~ \\
$2$&$1.35$&$0.89$&$1.42$&$-1.0$&$-2.1$&$11.2$&$4.2$ \\
\hline \hline
\end{tabular}
\end{table}

 \begin{figure}[tbp] 
 \centering
\includegraphics[scale=0.54]{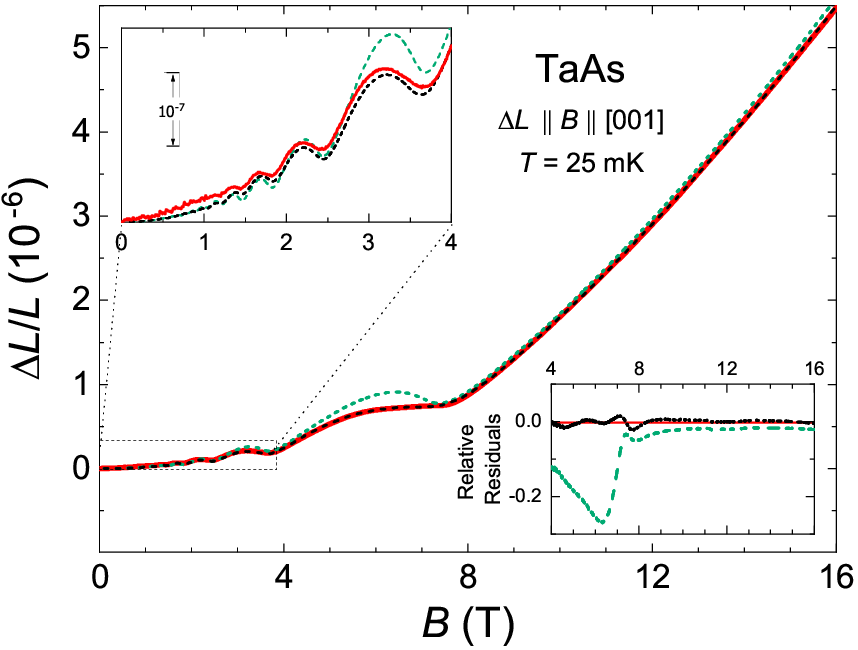}
\caption{\label{fig5}  \textbf{Comparison of the calculated magnetostriction for the case of $F_{W2}=1.35$ T with the experimental data  for TaAs.}
The solid red line shows the $c$-axis magnetostriction $\Delta L/L$ of TaAs measured at $T=25$ mK and $B\parallel c$. The dashed green and black lines have the same meaning as in Fig.~\ref{fig3}, but they are calculated for set 2 in Table I. The upper inset is a zoom into the low-field region, and the lower inset  compares the appropriate relative residuals, showing a very good agreement in the entire field range when $\zeta(B)$ is considered (dashed black line).
 } \end{figure}   

Interestingly, if one decreases the value of the parameter  $F_{W2}$, even better fits of the calculated magnetostriction to the experimental data can be obtained, and the best fit is reached at $F_{W2}\approx 1.35$ T (the dashed black line in Fig.~\ref{fig5}). Thus, we conclude that the two sets of the parameters are worth considering (Table I). The first set ($F_{W2}=5$ T) is completely consistent with the Fermi-surface calculations of Ref.~\cite{Arnold}, whereas the second set with  $F_{W2}=1.35$ T provides the best fit of the theoretical curve to our experimental data on the magnetostriction. The appropriate $\Lambda_{i}^c$ for the second set are presented in Table II, and in contrast to set 1, the analysis of $\beta$, $d\beta/d\zeta$, and $\nu_h$ obtained for set 2 reveals that the dispersion of the holes should essentially deviate from the parabolic law. However, apart from the problem of choosing the value of $F_{W2}$, a comparison of all the data presented in Figs.~\ref{fig3} and \ref{fig5} lead to the conclusion that in the first approximation, one can neglect the dependence of the chemical potential on the magnetic field (see also Supplementary Note 7). In other words, the simplified approach, within which $\zeta$ is independent of $B$, is sufficiently well justified for describing the magnetostriction.

In numerous experiments (see review \cite{m-sh19} and references therein), the phase of quantum oscillations in topological semimetals was  measured to distinguish  between the Weyl (Dirac) fermions and trivial quasiparticles. Such investigations are based on the fact that in the case of the Weyl fermions, the nonzero Berry phase of the electron orbits in a magnetic field leads to a shift of the phase of the oscillations by $\pi$ as compared to the phase corresponding to the trivial electrons \cite{m-sh19,prl}. Obviously, in agreement with formulas of Supplementary Notes $2$ and $3$, this nontrivial phase shift should also occur for the oscillations in the magnetostriction. For the case of TaAs ($\Delta L$\,$\parallel B$\,$\parallel c$), the  insets in Figs.~\ref{fig3} and \ref{fig5} clearly demonstrate that the phase of the oscillations calculated with formulas for the Weyl electrons really coincides with the experimental one. We emphasize that if the nontrivial phase shift were absent in the measured oscillatory magnetostrictions, the theoretical and experimental curves would be mutually displaced in phase like the red and black lines in Fig.~\ref{fig1}. Therefore, the coincidence of the phases, among other manifestations of the Weyl electrons in our magnetostriction measurements, also proves their existence in TaAs.

\section{Discussion}

Above we found the two sets of the parameters  for which the magnetostriction measured along the $c$ axis can be well described theoretically. To choose between these two set, we have measured  $\Delta L/L$ along the $a$ axis with the magnetic field still aligned with the $[001]$ direction. In this case, only the values of $a_{W1}$, $a_{W2}$, and $c_h$ can  change due to a change of the constants $\Lambda_{W1}$, $\Lambda_{W2}$, and $\Lambda_{h}$. All the other parameters determining this relative length change should remain the same as in the case of the $c$-axis magnetostriction.
Note that the knowledge of the constants $\Lambda_{i}$ for the magnetostriction measured not only along the $c$ axis but also along the $a$ direction makes it possible to find the constants of the deformation potential that describe effects of strains on the band structure of TaAs (Supplementary Notes 1 and 8).

\begin{figure}
	\begin{center}
\includegraphics[scale=0.54]{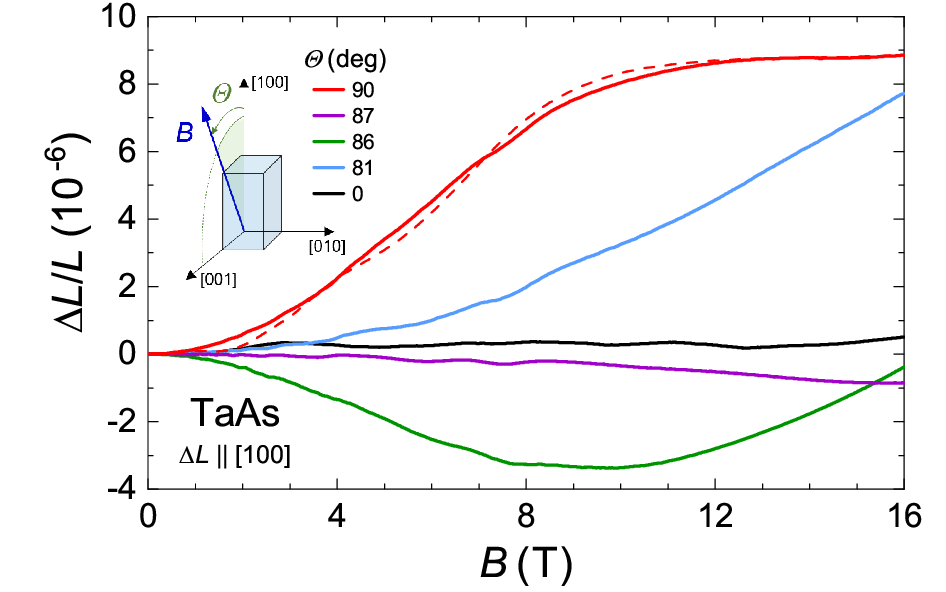}
\caption{\label{fig6} \textbf{Angle-dependent magnetostriction of TaAs measured along the $a$ axis.} Magnetic-field dependence of the relative length change $\Delta L/L$ of TaAs (sample 2) is measured along the [100] direction at 25\,mK for various angles ${\it \Theta}$ between the direction of $B$ and the $a$ axis.  The magnetic field lies in the plane (010). Note the exceptional changes of $\Delta L/L$ over the small-angle range near ${\it \Theta}$\,=\,90$^\circ$. (The corresponding results for ${\it \Theta}$\,$\leq$\,45$^\circ$ are shown in Supplementary Fig. 3). The dashed red line shows the $a$-axis magnetostriction that is calculated at $T$\,=\,0 for the field aligned with the $c$ direction, using $F_{W1}=7.2$ T, $F_{W1}=1.35$ T,  $\gamma_{W1}$\,=\,0.1 and  $\gamma_{W2}$\,=0.2 (Supplementary Note 8).}
	\end{center}
\end{figure}

\begin{figure}[tbp] 
 \centering  \vspace{+9 pt}
\includegraphics[scale=0.54]{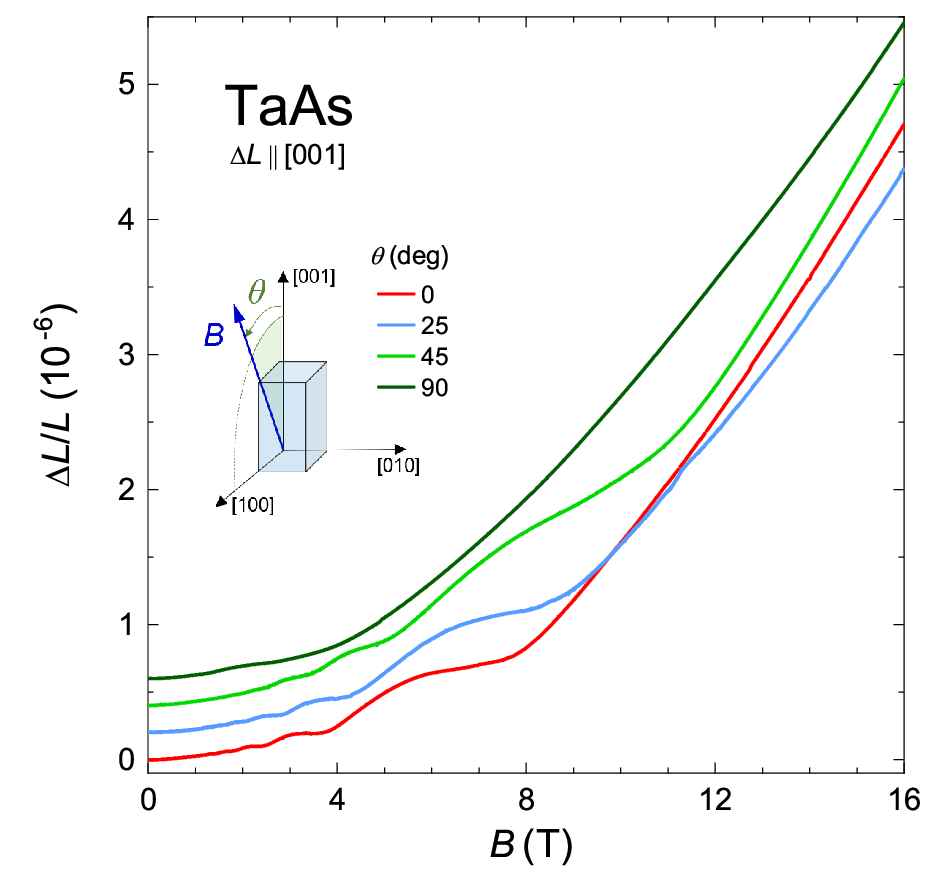}
\caption{\label{fig7} \textbf{Angle-dependent magnetostriction of TaAs measured along the $c$ axis.} Field-induced relative length change of TaAs (sample 3) along the [001] direction is measured at $T$\,=\,4.2\,K and at various angles $\theta$ between the direction of $B$ and the $c$ axis. The magnetic field lies in the plane (010). For clarity, the different $\Delta L/L$ curves at $\theta$\,$>$\,0 were shifted subsequently by $0.2\times10^{-6}$.
}
\end{figure}

Figure \ref{fig6} shows the magnetostriction measured along the $a$ axis at $T$\,=\,25\,mK. We note that a field-induced length change is small ($\sim$\,0.5\,$\times$\,10$^{-6}$ at $B$\,=\,16\,T) in magnetic fields applied along the dilatation direction (black curve). In addition, there is a complex $B$ dependence of $\Delta L/L$ in the entire field range.  However, when the dilatometer is rotated by the angle ${\it \Theta}$\,=\,90$^{\circ}$ to the desired sample orientation $B$\,$\parallel$\,$c$, the  $a$-axis magnetostriction exhibits a substantial enhancement with a behavior close to the $B^2$ law  between 0.5 and 5\,T. This behavior is followed by  a tendency to the saturation at about $9\times 10^{-6}$ above 12\,T (red curve). Another remarkable feature of the  $a$-axis magnetostriction is its high sensitivity to small deviations of the applied field from the [001] direction. Such  deviations  cause immense changes in the magnetostriction from large positive to large negative values. For example, the violet curve at  ${\it \Theta}$\,=\,87$^\circ$ illustrates the field-induced contraction of TaAs that is about $-$1.0\,$\times$\,10$^{-6}$ at $B$\,=\,16\,T. Moreover, a drastic change of the negative magnetostriction occurs when the magnetic field is just marginally  tilted further from the [001] direction (${\it \Theta}$\,=\,86$^\circ$, green curve). At ${\it \Theta}$\,=\,81$^\circ$ (blue curve), we again observe a large expansion of TaAs. In contrast with this $a$-axis magnetostriction, the $B$-induced length changes along the $c$ axis do not exhibit any sensitivity to small deviations of $B$ from the $c$ axis.
In Fig. \ref{fig7}, we present the angle-dependent magnetostriction of TaAs measured along the [001] direction at 4.2\,K. It is seen that there are no qualitative changes in this  magnetostriction even at large deviation angles. (Now the tilt angle $\theta$ of $B$ is measured from the $c$ axis.)

Varying only the constants $\Lambda_{i}$, we have calculated the $a$-axis magnetostriction at $B$\,$\parallel$\,$c$ for both sets of the parameters with $F_{W2}=5$ T and $1.35$ T (Supplementary Note 8). A modification of the values of $\gamma_{W1}$ and  $\gamma_{W2}$ is also admitted since Figs.~\ref{fig2} and \ref{fig6} show the magnetostrictions for the different samples. For the set with $F_{W2}=5$ T, we have not been able to match well the theoretical curve with the experimental data (cf. Supplementary Fig.~9). For the second set, the theoretical curve can approximately reproduce these data  (the dashed red line in Fig.~\ref{fig6}) for  certain values of $\Lambda_i\equiv \Lambda_i^{\perp}$ (Table II) and for the increased $\gamma_i$ ($\gamma_{W1}$\,=\,0.1 and  $\gamma_{W2}$\,=0.2). Thus, the obtained results seem to argue in favor of set 2 with  $F_{W2}=1.35$ T. However, due to the extreme sensitivity of the $a$-axis magnetostriction to the field orientation relative to the [001] direction, its true $B$ dependence at $B$\,$\parallel$\,[001] may essentially differ from the experimental curve (${\it \Theta}=90^{\circ}$) presented in Fig. 6. Thus, a more elaborate approach to achieve a perfect magnetic-field orientation is required in order to reliably exclude the possibility of set 1. In Supplementary Note 8, we discuss a possible cause of the unusual high sensitivity of the $a$-axis magnetostriction to a small tilting of $B$ about the $c$ axis.

The field-induced shift of the chemical potential presented in Fig.~\ref{fig4} clearly demonstrates the essential redistribution of the charge carriers between the bands of the holes and the W1 electrons. Although we found above that this effect can be of minor  significance for the magnetostriction of TaAs, it can be relevant to understanding some other field-dependent properties of Weyl-semimetal-candidate materials. In particular, the charge-carrier  redistribution leads to a nonzero longitudinal magnetoresistance of a semimetal if the quasiparticles in its different bands have dissimilar mobilities, and this magnetoresistance can be negative even for trivial charge carriers. Indeed, if with increasing $B$, the trivial electrons of a  lower mobility are transferred to another band with higher mobility, the longitudinal conductivity of this material increases (see, e.g. Fig.\,S10 in Supplemental Material to  Ref.~\cite{juraszek}). Therefore, the redistribution should be considered before giving any arguments in favor of the chiral anomaly in strong magnetic fields. In the case of TaAs,  our study reveals that for $B\parallel c$, i.e., the most promising configuration for the chiral anomaly \cite{Ramshaw}, both the absolute value $|n_h(B)|$ of the negative hole density and the density of the W1 electrons increase in high magnetic fields (Fig. \ref{fig4}, inset). In this situation, one may expect that the longitudinal conductivity along the [001] axis will increase above about 8\,T for any relation between mobilities of the electrons and holes. Interestingly, the negative longitudinal magnetoresistance for $B\parallel c$ was really observed in TaAs  at 7.5\,T $<$\,$B$\,$<$\,25\,T in careful measurements that used the focused-ion-beam lithography to eliminate experimental artifacts due to electrical current inhomogeneities \cite{Ramshaw}.

In summary, our study of the magnetostriction along the main crystallographic directions of TaAs shows that this quantity can be an effective probe of the massless quasiparticles in Weyl semimetals, if the Weyl points lie near the Fermi level. This statement holds even though conventional charge carriers exist in a semimetal. In this situation, even in moderate magnetic fields, which are too weak to confine large groups of massive quasiparticles at their zeroth Landau levels, the magnetostriction contains a linear-in-field term that identifies the presence of relativistic fermions. Moreover, in this case, the experimental magnetic-field and temperature dependences of the magnetostriction, including their subtle details,  can be reproduced theoretically. Comprehensive dilatometric investigations of topological semimetals  also shed light on dependences of the Weyl points on an applied  stress and hence  predict how the appropriate quantum-oscillation frequencies will change under a uniform compression of the material. It is also worth noting that our theory is applicable to Dirac semimetal.
Therefore, in a broader perspective, detection of relativistic fermions in candidate topological materials with the magnetostriction can set the stage for their further investigations, including electronic applications.

\section{Methods}

\textbf{Crystal synthesis.} TaAs single crystals were synthesized with the chemical vapor transport method following the procedure described elsewhere \cite{A,B,C}. The chemical composition of the crystals was examined by electron-probe microanalysis with energy-dispersive x-ray spectroscopy. The ratio of 1:0.99 between Ta and As was found,  indicating the correct stoichiometric chemical composition. The  body centered tetragonal structure (space group \textit{I}4$_1$\textit{md}, No.\,109) of TaAs single crystals was confirmed by room-temperature x-ray diffraction. No other phases were detected, and the lattice parameters $a$\,=\,3.4348\,\AA\ and $c$\,=\,11.6412\,\AA\ are in good agreement with the literature values \cite{A,B,C}. The crystal orientation was determined by Laue diffraction (cf. Supplementary Fig. 1,right).

\textbf{Sample characterization.} All samples used in this study were obtained from the same growth batch. The residual-resistivity ratio of typical sample RRR\,=\,{$\rho_{300\,\rm{K}}$}/$\rho_{2\,\rm{K}}$\,$\simeq$\,12 was determined from a zero-field resistance measurement along the $c$ axis. The transverse magnetoresistance MR of about 39 600\%  was measured at 9\,T and 2\,K (Supplementary Fig. 2). These parameters are in good agreement with the published literature values, and hence point at good quality of our TaAs single crystals.

\textbf{Magnetostriction measurements.}  The angle-dependent field-induced length change was measured with a commercial  capacitance dilatometer which enables a length resolution of 0.02\,$\AA$ \cite{Kuchler2}. The magnetostriction of  the three  rectangular TaAs samples having a length of approximately 1.7\,mm was studied along the [100] and [001] directions. We performed the magnetostriction measurements down to 25 mK in a dilution refrigerator (Kelvinox\,400 HA, Oxford Instruments) inserted into a superconducting magnet for fields up to 16\,T. A field-sweep rate as small as 0.5\,mTs$^{-1}$ was used, and the highest temperature of our experiments was 4.2\,K.

The capacitive dilatometer cell is compact enough to be mounted on an attocube rotator, and thus enables the study of the field-induced length changes as a function of the tilted angle. However, since TaAs single crystals with a typical cross-section of about 0.7\,$\times$\,0.9\,mm$^{2}$ were mounted between two capacitor plates with the diameter of 20 mm, the samples cannot be oriented perfectly. The limited accuracy of orientation of the sample surfaces with respect to the dilatometer setup can cause the maximum error of $5^{\circ}$.

\textbf{Numerical calculations.} The $B$-dependences of the magnetostriction for TaAs have been numerically calculated, using our own code elaborated with formulas of Supplementary Notes $1-4$ and results of Refs.~\cite{harris,jones,johansson}.

\textbf{Data Availability.} The data that support the plots within the main manuscript or the supplement and other findings of this study are available from the corresponding authors upon a reasonable request.

\textbf{Additional Information.} Supplementary information accompanies this paper (see pages 11-25).

\section{Acknowledgements}

This work was supported by the Polish National Science Centre (Project No. 2016/21/B/ST3/02361).




\begin{titlepage}
	\begin{center}
		\vspace*{2.4cm}
		
		\huge Supplementary Information
		
		\vspace{0.5cm}
		\Huge	Detection of relativistic fermions in Weyl semimetal TaAs by magnetostriction measurements
		
		\vspace{1.5cm}
		
		\Large	Tomasz Cichorek, \L ukasz Bochenek, and Jaros\l aw Juraszek
		\vspace{0.2cm}
		
		\large \textit{Institute of Low Temperature and Structure Research,
			\\
			Polish Academy of Sciences, 50-422 Wroc\l aw, Poland}
		
		\vspace{1cm}
		
		\Large	Yuriy V. Sharlai and Grigorii P. Mikitik
		\vspace{0.2cm}
		
		\large \textit{B. Verkin Institute for Low Temperature Physics and Engineering,
			\\
			Ukrainian Academy of Sciences, Kharkov 61103, Ukraine}
		\vfill

		\vspace{0.8cm}

	\end{center}
\end{titlepage}

\clearpage
\addtocounter{figure}{-7}
\addtocounter{table}{-2}
\addtocounter{page}{10}
\renewcommand{\figurename}{Supplementary Fig.}
\renewcommand{\tablename}{Supplementary Table.}

\begin{figure*}[htb]
	\begin{center}
		\includegraphics[height=45mm]{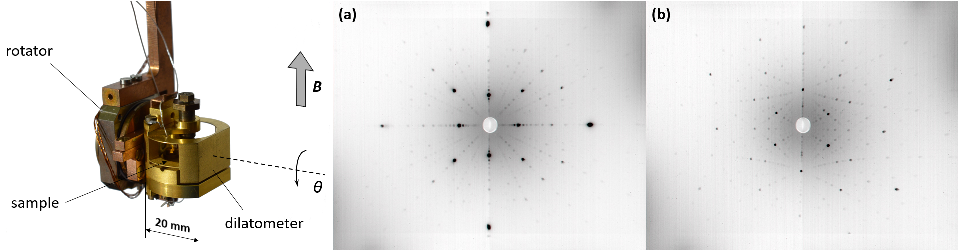}
		\caption{Left: Photograph  of  the commercial dilatometer cell \cite{Kuchler2} mounted  on  a  cold  finger  of  a  dilution refrigerator. This setup is equipped with a piezoelectric rotator for angle-dependent magnetostriction. Right: Back-reflection Laue photographs of the TaAs single crystal (sample\,1). Figure (a) shows the diffractogram along the $c$ axis, whereas (b) shows the diffractogram along the $a$ axis.}
		\label{fig:FigSI2}
	\end{center}
\end{figure*}

\begin{figure*}[htb]
	\begin{center}
		\includegraphics[width=172mm]{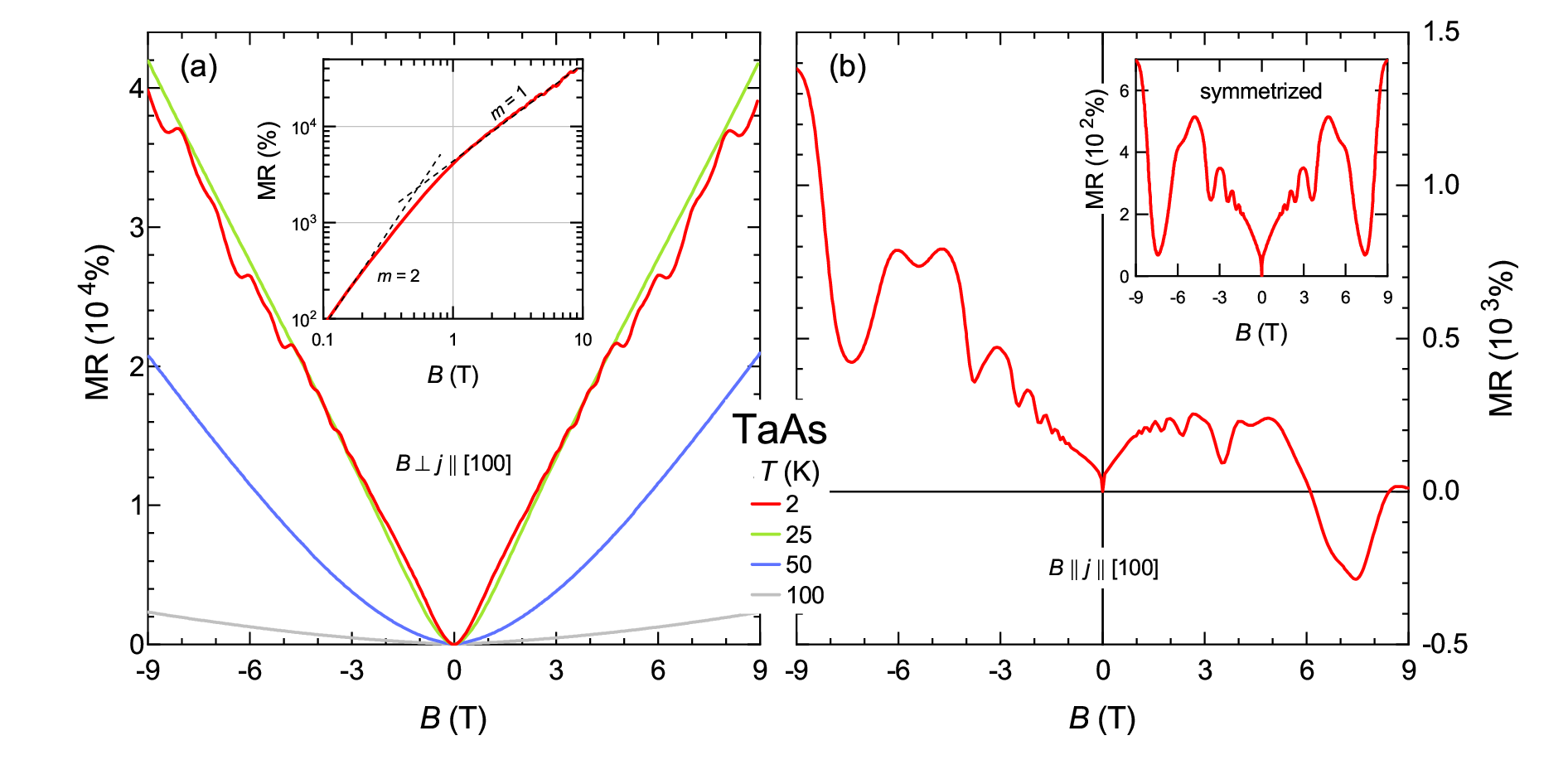}
		\caption{Transverse (a) and longitudinal (b) magnetoresistance of TaAs measured along the $a$ axis. No off-diagonal components were subtracted from the MR data shown in main panels.}
		\label{fig:FigSI3}
	\end{center}
\end{figure*}


\begin{figure*}[htb]
	\begin{center}
		 \includegraphics[width=\linewidth,keepaspectratio=true]{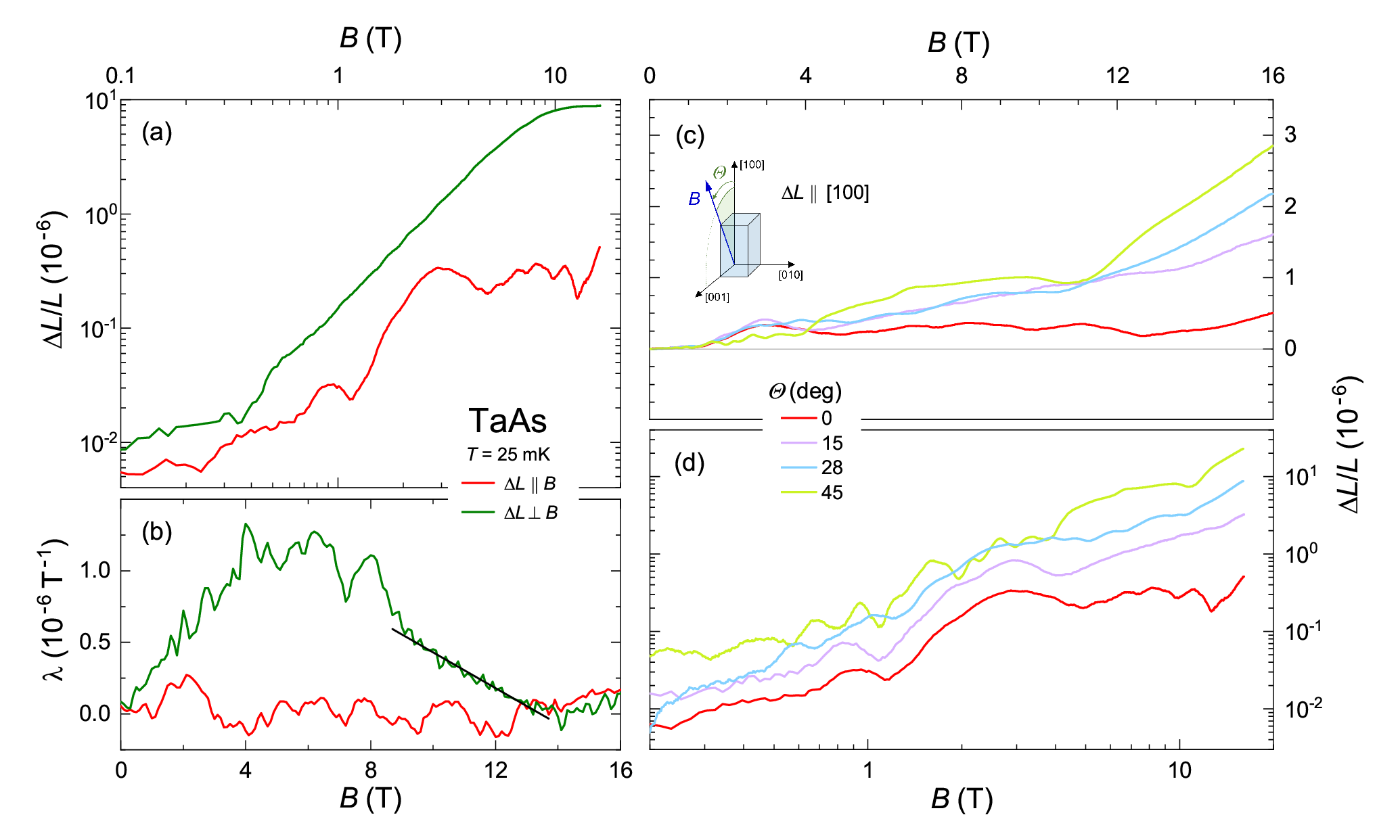}
		\caption{The $a$-axis magnetostriction of TaAs measured within the (010) plane at 25\,mK. (a) Magnetic-field dependence of the relative length change $\Delta L/L$ measured along the [100] direction  both in the parallel (red) and perpendicular (green) configuration. At 16\,T, the $\Delta L$\,$\parallel$\,$B$ and $\Delta L$\,$\perp$\,$B$ data  differ by more than one order of magnitude. (b) Magnetostriction coefficient $\lambda$ derived from the longitudinal and transverse measurements. The black straight line with the slope $-12.5\times 10^{-8}$ marks the range where $\lambda$ is a linear function of $B$. (c) Complex dependences of the magnetostriction at small tilt angles $\Theta$\,$\le$\,45$^\circ$. Note a small longitudinal magnetostriction in the entire field range (red). (d) Log-log plot of the experiential results shown in (c). Curves are subsequently shifted by a factor of 3 for clarity.}
		\label{fig:Fig1}
	\end{center}
\vspace{70mm}
\end{figure*}

\clearpage
		
\twocolumngrid
\section{Supplementary Note $1$: Formula for the magnetostriction}

We now derive the general formula for the magnetostriction [formula (1) in the main text], assuming, for definiteness,  the tetragonal symmetry of a crystal (i.e., the symmetry of TaAs). This formula can be obtained by a minimization of the $\Omega$ potential for the deformed crystal placed in the external magnetic field $H$ with respect to the strain tensor $u_{ik}$,
\begin{eqnarray}\label{1a}
\Omega&=&\frac{1}{2}C_{11}(u_{xx}^2+u_{yy}^2)+\frac{1}{2}C_{33}u_{zz}^2
\nonumber \\ &+&C_{13}(u_{xx}+u_{yy})u_{zz}+C_{12}u_{xx}u_{yy}\nonumber \\&+&2C_{66}u_{xy}^2 + 2C_{44}(u_{xz}^2+u_{yz}^2)\nonumber \\&+&\Delta \Omega_{\rm el}(u_{ik},B)- \Delta \Omega_{\rm el}(u_{ik},0),
\end{eqnarray}
where $C_{mn}$ are the elastic moduli of the crystal \cite{LL-sl}, $B=\mu_0H$ is the magnetic induction in the sample,  $\Delta \Omega_{\rm el}(u_{kl},B)\equiv
\Omega_{\rm el}(u_{kl},B)-\Omega_{\rm el}(0,B)$, and $\Omega_{\rm el}(u_{kl},B)=\sum_{i=1}^N \Omega_{i}(u_{kl},B)$ is the deformation-dependent part of the electron $\Omega$ potential for the crystal with $N$ pockets of charge carriers. This part depends on the deformation, the magnetic induction  $B$, and also on the temperature $T$ and the chemical potential $\zeta$ (we do not indicate explicitly $T$ and $\zeta$ here). The first six terms in Eq.~(\ref{1a}) give the total elastic energy of a deformation. This energy is partly produced by  $\Delta \Omega_{\rm el}(u_{kl},0)$, and hence the difference of the elastic terms and $\Delta \Omega_{\rm el}(u_{kl},0)$ is the elastic energy that is not associated with the electron pockets under study. The term $\Delta \Omega_{\rm el}(u_{kl},B)$ describes the total change in the $\Omega$ potential of these pockets in the magnetic field under the deformation. As in Ref.~\cite{lif,Keyes,Mi,m-sh15}, we assume that an elastic deformation shifts the electron bands as a whole and does not change their shape.  This rigid-band approximation commonly is well justified for sufficiently small pockets (see below). In particular, this approximation seems to be valid for TaAs family of the semimetals, see Figs.~6e and 6f in Ref.~\cite{reis}. In this rigid-band approximation, a shift $\Delta\varepsilon_i$ of the $i$th electron energy band $\varepsilon_i({\bf p})$ under the deformation coincides with the shift of its edge and is proportional to the strain tensor $u_{kl}$, $\Delta\varepsilon_i=\sum_{k,l}\lambda_{kl}^{(i)}u_{kl}$, where ${\bf p}$ is the electron quasimomentum, and the constants $\lambda_{kl}^{(i)}$ are the so-called deformation potential of the $i$th band \cite{abr}. Therefore, in the minimization of $\Omega$ potential, one can use the following relation:
\[
\frac{\partial\Omega_{i}(u_{kl},B)}{\partial u_{kl}}= -\frac{\partial\Omega_{i}(\zeta,B)}{\partial \zeta} \frac{\partial\Delta\varepsilon_i}{\partial u_{kl}}=\lambda_{kl}n_i(B),
\]
where $n_i(B)=-\partial\Omega_{i}(\zeta,B)/\partial \zeta$ is the density of charge carriers of the pocket $i$ in the magnetic field. The minimization gives the set of the equations in the tensor  $u_{kl}$, which defines the magnetostriction, i.e., the deformation of the crystal in the magnetic field,
\begin{eqnarray*}
C_{33}u_{zz}&+&C_{13}(u_{xx}+u_{yy})=-\sum_i\lambda_{zz}^{(i)}[n_i(B)-n_i(0)], \\
C_{11}u_{xx}&+&C_{13}u_{zz}+C_{12}u_{yy}=-\sum_i\lambda_{xx}^{(i)}[n_i(B)-n_i(0)], \\
C_{11}u_{yy}&+&C_{13}u_{zz}+C_{12}u_{xx}=-\sum_i\lambda_{yy}^{(i)}[n_i(B)-n_i(0)],\\
4C_{66}u_{xy}&=&-\sum_i\lambda_{xy}^{(i)}[n_i(B)-n_i(0)], \\
4C_{44}u_{xz}&=&-\sum_i\lambda_{xz}^{(i)}[n_i(B)-n_i(0)], \\
4C_{44}u_{yz}&=&-\sum_i\lambda_{yz}^{(i)}[n_i(B)-n_i(0)],
\end{eqnarray*}
where $i=1,\dots, N$ usually runs several groups of the equivalent Fermi pockets. (We call pockets equivalent if, in absence of the magnetic field, they transform into each other under symmetry operations.)  It follows from symmetry considerations that inside each of the groups, the parameters $\lambda_{zz}^{(i)}$ have one and the same value, the  $\lambda_{xx}^{(i)}$ and $\lambda_{yy}^{(i)}$ take on only the two values $\lambda_{xx}^{a1}$ and $\lambda_{xx}^{a2}$ with  $\lambda_{yy}^{a1}=\lambda_{xx}^{a2}$, $\lambda_{yy}^{a2}=\lambda_{xx}^{a1}$, whereas
the off-diagonal $\lambda_{lz}^{(i)}$ (with $l=x, y$) run the four values $\pm\lambda_{lz}^{a1}$, $\pm\lambda_{lz}^{a2}$, and $\lambda_{xy}^{(i)}$ is equal to $\pm\lambda_{xy}$.
Solving this set of the equations, we obtain,
\begin{eqnarray}\label{2a}
u_{zz}&=&\sum_i\Lambda_i[n_i(B)-n_i(0)], \\
\Lambda_i&=&-\frac{\lambda_{zz}^{(i)}(C_{11}+C_{12}) -2C_{13}\bar\lambda_{xx}^{(i)}}{C_{33}(C_{11}+C_{12})-2(C_{13})^2} \equiv \Lambda_i^c.
\label{3a}
\end{eqnarray}
The same formula (\ref{2a}) describes $(u_{xx}+u_{yy})/2$ but with the other $\Lambda_i$,
\begin{eqnarray}\label{4a}
\Lambda_i=-\frac{\bar\lambda_{xx}^{(i)}C_{33}- \lambda_{zz}^{(i)}C_{13}}{C_{33}(C_{11}+C_{12})-2(C_{13})^2}\equiv \Lambda_i^{\perp}.
\end{eqnarray}
In Eqs.~(\ref{3a}) and (\ref{4a}), $\bar\lambda_{xx}^{(i)}\equiv  (\lambda_{xx}^{a1}+\lambda_{xx}^{a2})/2$ for all $i$ belonging to the appropriate group of the equivalent pockets. It is significant that $\Lambda_i^c$ and $\Lambda_i^{\perp}$ have one and the same values inside each of these groups. On the other hand, the coefficients $\Lambda_i$ in formula (\ref{2a}), which also describes $u_{xx}$ and  $u_{yy}$, take on the two different values for the equivalent Fermi pockets:
\begin{eqnarray*}
\Lambda_i^{a1}=\Lambda_i^{\perp}-\frac{\lambda_{xx}^{a1}-\lambda_{xx}^{a2}}{2 (C_{11}-C_{12})},\ \ \  \Lambda_i^{a2}=\Lambda_i^{\perp}+\frac{\lambda_{xx}^{a1}-\lambda_{xx}^{a2}}{2 (C_{11}-C_{12})},
\end{eqnarray*}
where $a1$ and $a2$ mark such pockets lying near the two different reflection planes.

When the magnetic field is along the $z$ axis, all $n_i(B)$ are one and the same for the equivalent Fermi surfaces, the contributions of the second terms in $\Lambda_i^{a1}$ and $\Lambda_i^{a1}$ to the sums over $i$ cancel each other out, and we obtain from the above set of the equations that $u_{xx}=u_{yy}$, and $u_{xy}=u_{zx}=u_{zy}=0$. In this situation, the $u_{zz}$ is the elongation $\Delta L/L$ of the crystal along the direction $[001]$, whereas $u_{xx}$ is the elongation in the direction $[100]$. Formulas (\ref{3a}) and (\ref{4a}) show that the constants $\Lambda_i^c$ and $\Lambda_i^{\perp}$ extracted from the measurements of $u_{zz}$ and $u_{xx}$ enable one to find the constants $\lambda_{zz}^{(i)}$, $\bar\lambda_{xx}^{(i)}$ of the deformation potential if the the elastic moduli $C_{mn}$ are known. In particular, these moduli for TaAs are given in Ref.~\cite{buckeridge}, and for this semimetal, formulas (\ref{3a}) and (\ref{4a}) yield the following equations:
\begin{eqnarray} \label{5a}
\Lambda_i^c&\approx& (-8.75\lambda_{zz}^{(i)}+4.94\bar\lambda_{xx}^{(i)})\times 10^{-25} {\rm cm}^{3},\\
\Lambda_i^{\perp}&\approx& (2.48\lambda_{zz}^{(i)}-4.67\bar\lambda_{xx}^{(i)})\times 10^{-25} {\rm cm}^{3}, \nonumber
\end{eqnarray}
where $\lambda_{zz}^{(i)}$ and $\bar\lambda_{xx}^{(i)}$ are expressed in eV. Note that formula (\ref{2a}) was used in a number of papers \cite{Keyes,Mi,m-sh15}. (In Refs.~\cite{m-sh15}, corrections to Eq.~(\ref{2a}) were also taken into account when the chemical potential is close to the edge of a Landau subband.) For definiteness, we shall discuss only $u_{zz}$ in the subsequent Supplementary Notes 2-7.

When the magnetic field deviates from the $z$ axis, the densities $n_i(B)$ for different equivalent pockets generally do not coincides, and in this case, $u_{xx}\neq u_{yy}$, and the off-diagonal $u_{ik}\neq 0$. The latter means that the diagonalization of the tensor $u_{kl}$ gives its principal axes that differ from the axes $x$, $y$, $z$, and hence, the deformation of the crystal in such a tilted magnetic field is a superposition of its elongations (contractions) along the three directions different from the crystallographic axes. Such elongations change the shape of the sample. For example, if the sample at $B=0$ is a rectangular parallelepiped with the axes $x$, $y$, $z$, it becomes an oblique parallelepiped. When $B$ deviates from the $z$ axis in the direction of the $x$ axis, the cross sections of this parallelepiped by the planes $y=$const.\ are  parallelograms, the angle of which differs from $\pi/2$ by a small  amount of the order of $u_{xz}$. If we define $\Delta L$ as the change in the distance between the parallel surfaces of the parallelepiped, the relative elongation $\Delta L/L$ is still given by $u_{zz}$ (or $u_{xx}$). Another consequence of the tilt of $B$ is that the different values of the factors $\Lambda_i^{a1}$ and $\Lambda_i^{a2}$ for the equivalents pockets manifest themselves in the expression for $u_{xx}$, whereas the factor $\Lambda_i^c$ is still one and the same for these pockets. This fact may lead to essentially different dependences of $u_{xx}$ and $u_{zz}$ on the direction of $B$ (Supplementary Note 8).

Finally, let us comment on the applicability of the rigid-band approximation, which  has been used above. In general case, the shift $\Delta\varepsilon_i({\bf p})=\sum_{k,l}\lambda_{kl}^{(i)}({\bf p})u_{kl}$ of the $i$th electron energy band $\varepsilon_i({\bf p})$ under the deformation depends on ${\bf p}$ due to the ${\bf p}$ dependence of $\lambda_{kl}^{(i)}({\bf p})$ \cite{abr}. The deformation potential $\lambda_{kl}^{(i)}({\bf p})$ essentially changes on the scale of the order of the size of the Brillouin zone, and so one can consider $\lambda_{kl}^{(i)}$ as constant for small pockets of the Fermi surface. However, a caution is required when a band of the trivial electrons is separated from a valence band by a  sufficiently small gap $\Delta$. In this case, the effective electron masses of the band are of the order of $\Delta/V^2$ where $V$ is a typical interband matrix element of the velocity operator ($V\sim 10^5\div 10^6$ m/s). If the small gap $\Delta$ substantially depends on a deformation, the rigid-band approximation can fail for a small pocket of this band since the effective masses, and hence the shape of the band, noticeably change with the deformation. Another  situation occurs for the Weyl nodes. These nodes are topologically protected, a small deformation cannot create a gap in the spectrum, and so one may expect the applicability of the rigid-band approximation to the case of the Weyl pockets.
		
\section{Supplementary Note $2$: Magnetostriction produced by  Weyl electrons in noncentrosymmetric crystals}

As was shown in Sec.~3.3 of Ref.~\cite{m-sh19}, for a {\it  noncentrosymmetric} crystal, a contribution of a Fermi pocket of Weyl quasiparticles to any thermodynamic quantity (including the magnetostriction) can be calculated as half of the appropriate contribution of the Dirac pocket with the same electron dispersion. Using formulas for the electron spectrum in the magnetic field in the vicinity of a Dirac point \cite{m-sh19,m-sh}, the $\Omega$ potential and hence the density $n(B)$ of the Weyl electrons can be found. In particular, we obtain at zero temperature,
\begin{eqnarray}\label{6a}
n(\zeta-\varepsilon_d,B)
=n(\zeta- \varepsilon_d,0)\Big(\frac{3}{4u}+\!\frac{3}{2u^{3/2}} \!\! \sum_{n=1}^{n=[u]}\!\!\sqrt{u-n}\Big),~~
\end{eqnarray}
where $n(\zeta-\varepsilon_d,B)$ is the density of the Weyl quasiparticles in the magnetic field $H=B/\mu_0$ at a given chemical potential $\zeta$; the density in absence of the magnetic field, $n(\zeta-\varepsilon_d,0)$, is assumed to be positive in the case of the electrons ($\zeta-\varepsilon_d>0$) and negative for the holes ($\zeta-\varepsilon_d<0$); $\varepsilon_d$ is the energy of the Weyl point; $[u]$ means the integer part of $u\equiv F/B$; the frequency of the quantum oscillations
\begin{equation}\label{7a}
F= \frac{S_{\rm max}}{2\pi e\hbar}
\end{equation}
also defines the boundary of the ultra-quantum region  at which these oscillations disappear, and $S_{\rm max}$ is the maximal cross sectional area of the Fermi surface of the Weyl quasiparticles at a given direction of the magnetic field. This area is proportional to $(\zeta-\varepsilon_d)^2$. Of course, formula (\ref{6a}) is applicable to the case of the Dirac electrons, too. We also point out the two useful representations of the sum in Eq.~(\ref{6a}),
\begin{eqnarray}\label{8a}
\sum_{n=1}^{n=[u]}\sqrt{u-n}&=&\zeta(-1/2,\{ u\})-\zeta(-1/2,u),\\
\sum_{n=1}^{n=[u]}\sqrt{u-n}&=&-{\rm Im}\{\zeta(-1/2,1-u-0i)\},
\label{9a}
\end{eqnarray}
where $\zeta(-1/2,u)$ is the Hurwitz zeta function \cite{batem}, and $\{ u\}\equiv u-[u]$ is the fractional part of $u$.

We emphasize that formula (\ref{6a}) is valid for the most general form of the dispersion laws $\varepsilon_{c}({\bf p})$ and $\varepsilon_{v}({\bf p})$ describing the quasiparticles in the vicinity of the Weyl point \cite{m-sh19},
\begin{eqnarray*}
	\varepsilon_{c,v}({\bf p})&=&\varepsilon_d+{\bf a}\cdot{\bf p}\pm E({\bf p}), \\
	E({\bf p})&=&[b_{11}p_1^2+b_{22}p_2^2+b_{33}p_3^2]^{1/2},
\end{eqnarray*}
where ${\bf a}=(a_1,a_2,a_3)$ and $b_{ii}$ are the constant parameters, with  $a_i$ specifying the so-called tilt of the spectrum. For reference, let us also write the expressions for $n(\zeta- \varepsilon_d,0)$ and $S_{\rm max}$ in terms of these parameters,
\begin{eqnarray*}
	n(\zeta- \varepsilon_d,0)&=&\frac{(\zeta-\varepsilon_d)^3}{3 \pi^2 \hbar^3 (b_{11}b_{22}b_{33})^{1/2}(1-\tilde a^2)^2}, \\
	S_{\rm max}&=&\frac{\pi(\zeta-\varepsilon_d)^2}{(1-\tilde a^2)R_n^{1/2}},
\end{eqnarray*}
where $\tilde a^2\equiv \tilde a_1^2+\tilde a_2^2+\tilde a_3^2$, $\tilde a_i\equiv a_i/\sqrt{b_{ii}}$,
\begin{eqnarray*}
	R_n&=&\sum_{i,j=1}^{3}\kappa^{ij}n_in_j \\
	\kappa^{ij}\!\!&=&\frac{b_{11}b_{22}b_{33}}{(b_{ii}b_{jj})^{1/2}} \left[(1-\tilde a^2)\delta_{ij}+\tilde a_i\tilde a_j\right], \end{eqnarray*}
$\delta_{ij}$ is the Kronecker symbol, and ${\bf n}=(n_1,n_2,n_3)$ is the unit vector along the magnetic field.

Formulas (\ref{2a}) and (\ref{6a}) describe the magnetostriction $u_{zz}$ produced by the Weyl electrons at zero temperature $T$ and in absence of the electron scattering by impurities.
It was demonstrated by Shoenberg \cite{Sh} that the effect of the impurities on the oscillating part of the magnetization can be taken into account by replacing the Fermi energy $E_F$ (i.e. $\zeta$ at $T=0$) by $E_F+i\Gamma$ where the parameter $\Gamma=\pi T_D$ characterizes the electron scattering by the impurities, and $T_D$ is the so-call Dingle temperature. We shall use this replacement both for the oscillating and for the smooth parts of the magnetostriction. In other words, to take into account the scattering, the parameter $u+0i$ in formula (\ref{9a}) is replaced by $u(1+i\gamma)^2$ where $\gamma\equiv \Gamma/(\zeta-\varepsilon_d)=\pi t_D$ is the dimensionless  characteristic of the scattering, and $t_D$ is the dimensionless Dingle temperature. As to the magnetostriction at $T>0$, it can be obtained as follows:
\begin{equation}\label{10a}
u_{zz}(\zeta,T,B)=\int_{\!-\infty}^{\infty}\!\! d\varepsilon  \frac{u_{zz}(\varepsilon,0,B)}{4T\cosh^2[(\varepsilon-\zeta)/2T]}.
\end{equation}
Consider now various limiting cases.

\begin{figure}[tbp] 
	\centering  \vspace{+9 pt}
	\includegraphics[scale=0.47]{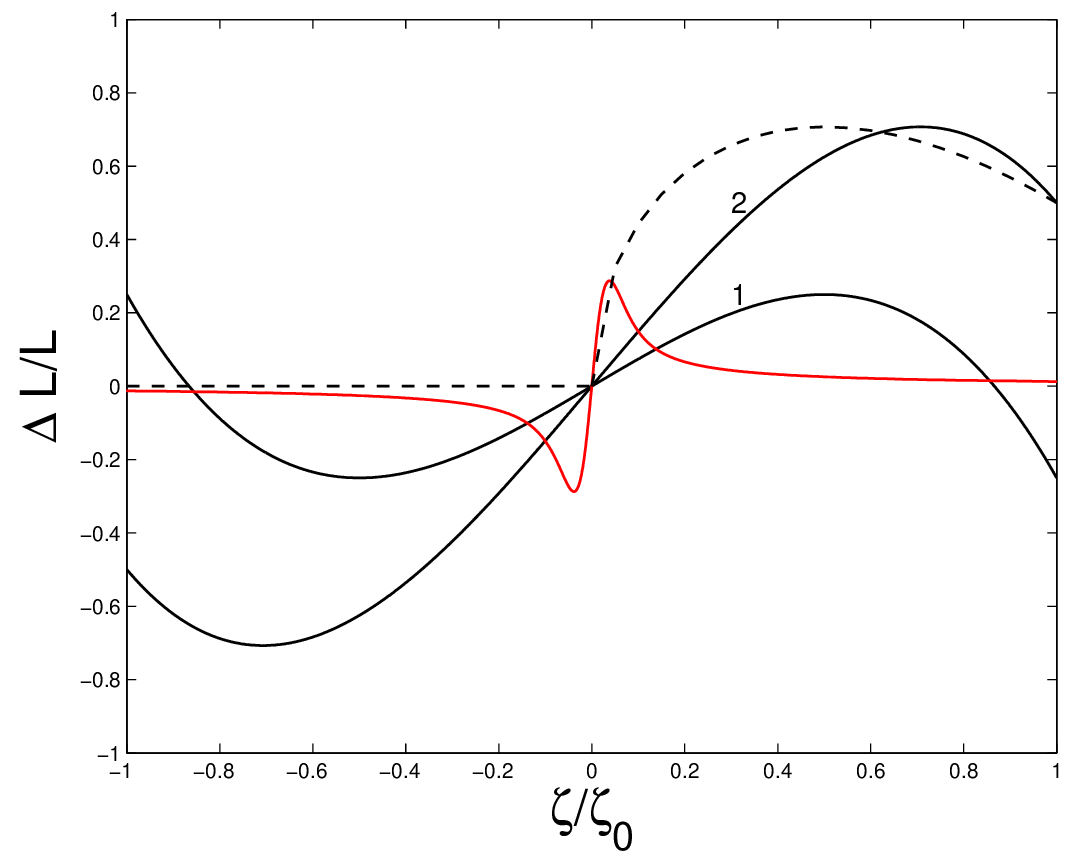}
	\caption{\label{fig1a}
		The magnetostriction $\Delta L/L=u_{zz}$ (in units of $\Lambda n(\zeta_0,0)$) produced by the Weyl electrons at $T=0$ versus the dimensionless chemical potential $\zeta/\zeta_0$ for $\mu_0H/F=2$ and $\mu_0H/F=1$ (the solid black lines, the numbers near these lines indicate the values of $\mu_0H/F$). The chemical potential $\zeta$ is measured from the $\varepsilon_d$, and $\zeta_0$ is its initial value. The solid red line shows the smooth part of the magnetostriction, $\bar u_{zz}$, enlarged in $10$ times for $\mu_0H/F=1/7$ and for the nonzero temperature $T/\zeta_0=0.02$. The dashed line depicts the dependence of the magnetostriction on the chemical potential $\zeta$ at $T=0$, $\mu_0H/F=2$ for the crystal with the electron spectrum (\ref{18a}) and $\delta=1/2$ (the chemical potential is now measured from $\varepsilon_0$). Note that this dependence, which vanishes below $\varepsilon_0$,  qualitatively differs from the dependence for the Weyl electrons.
} \end{figure}   

{\bf Ultra-quantum regime ($\mu_0H>F$).}
At high fields $\mu_0H>F$, the sum in equation (\ref{6a}) disappears since $u=F/B<1$ in this case. Then, with formula (\ref{2a}), we obtain  the following contribution of a Weyl pocket to the magnetostriction:
\begin{eqnarray}\label{11a}
u_{zz}=\Lambda n(\zeta- \varepsilon_d,0) \left(\frac{3B}{4F}-1\right)\equiv a+bB,
\end{eqnarray}
where the quantities $\Lambda$, $n(\zeta- \varepsilon_d,0)$, $F$ introduced above and $a\equiv -\Lambda n(\zeta- \varepsilon_d,0)$,  $b\equiv -0.75a/F$ refer to  this pocket. It is evident that the Weyl pocket produces the elongation $\Delta L/L$ that increases linearly with $B$ at the high magnetic fields.

Consider the dependence of $u_{zz}$ on the chemical potential $\zeta$ at a fix value of $B$ (Supplementary Fig.~\ref{fig1a}). Since $F\propto S_{\rm max}\propto (\zeta-\varepsilon_d)^2$ and $n(\zeta- \varepsilon_d,0)\propto (\zeta-\varepsilon_d)^3$, we arrive at
\begin{eqnarray*}
	u_{zz}=\Lambda n(\zeta_0- \varepsilon_d,0)\frac{(\zeta-\varepsilon_d)^3}{(\zeta_0-\varepsilon_d)^3} \left(\frac{3B}{4F}\frac{(\zeta_0- \varepsilon_d)^2}{(\zeta-\varepsilon_d)^2}-1\right),
\end{eqnarray*}
where $\zeta_0$ is the initial value of the chemical potential. Supplementary figure \ref{fig1a} shows that the magnetostriction changes its sign when $\zeta$ crosses the energy of the Weyl point $\varepsilon_d$.

{\bf Weak magnetic fields ($\mu_0H\ll F$).}
At $\mu_0H\ll F$, the ratio $u$ in Eq.~(\ref{6a}) is large, and the sum contains many terms. Using the representation (\ref{8a}) and the expansion of the Hurwitz zeta function $\zeta(-1/2,u)$ at $u\gg 1$ \cite{batem},
\begin{eqnarray}\label{12a}
\zeta(-1/2,u)\!\approx -\frac{2}{3}u^{3/2}+\frac{1}{2}u^{1/2} -\frac{1}{24u^{1/2}},
\end{eqnarray}
we find from formulas (\ref{2a}) and (\ref{6a}) that the magnetostriction $u_{zz}$ produced by a Weyl pocket comprises the smooth and oscillating parts in the weak-field range, $u_{zz}=\bar u_{zz} +u_{zz}^{\rm osc}$. The smooth part $\bar u_{zz}$ has the form
\begin{equation}\label{13a}
\bar u_{zz}= \Lambda n(\zeta-\varepsilon_d,0) \frac{B^2}{16F^2}= -\frac{b^2}{9a}B^2\equiv cB^2,
\end{equation}
and the oscillating part $u_{zz}^{\rm osc}$ results from the term $\zeta(-1/2,\{ u\})$ in Eq.~(\ref{8a}),
\begin{equation}\label{14a}
u_{zz}^{\rm osc}= \Lambda n(\zeta-\varepsilon_d,0) \frac{3\zeta(-1/2,\{ u\})}{2u^{3/2}}.
\end{equation}
The oscillating behavior of $u_{zz}^{\rm osc}$ with changing $u=F/B$ becomes evident from the formula \cite{batem},
\begin{eqnarray}\label{15a}
\zeta(-1/2,\{ u\})= \frac{1}{2\pi\sqrt{2}} \sum_{n=1}^{\infty}\frac{1}{n^{3/2}}\,\sin\!\!\left(2\pi nu-\frac{\pi}{4}\right).
\end{eqnarray}

Interestingly, with changing the chemical potential, the coefficient $c$ sharply changes it sign when $\zeta$ crosses the energy of the Weyl point (Supplementary Fig.~\ref{fig1a}),
\begin{eqnarray}\label{16a}
c(\zeta)=c(\zeta_0)\frac{(\zeta_0- \varepsilon_d)}{(\zeta-\varepsilon_d)},
\end{eqnarray}
where $\zeta_0$ is the initial value of $\zeta$. Note that $c(\zeta)$  diverges at $\zeta=\varepsilon_d$.

Formulas (\ref{11a}), (\ref{13a}), (\ref{14a}), and (\ref{16a}) are written at $T=0$. A nonzero temperature tends to suppress the oscillations of the magnetostriction, but it has effect on $\bar u_{zz}$ only in the region $|\zeta-\varepsilon_d|\lesssim T$. In this region, the divergence of $c(\zeta)$ near $\varepsilon_d$ is cut off  (Supplementary Fig.~\ref{fig1a}), and the maximal value of $c(\zeta)$ is of the order of $c(\zeta_0)(\zeta_0- \varepsilon_d)/T$.

\begin{figure}[t] 
	\centering  \vspace{+9 pt}
	\includegraphics[scale=1.]{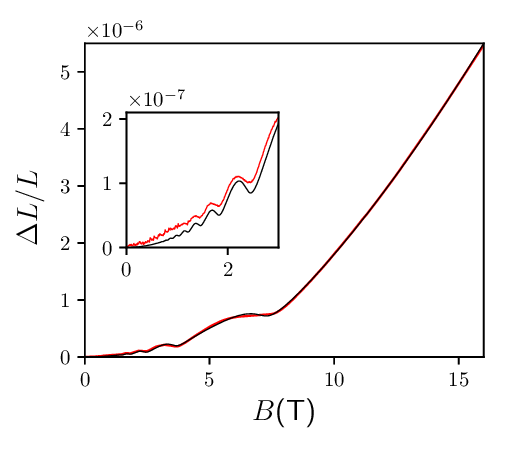}
	\caption{\label{fig2a} Comparison of the calculated magnetostriction with the experimental data  for TaAs.
		The red line shows the $c$-axis magnetostriction $\Delta L/L$ of the crystal TaAs measured at $T=25$ mK for the magnetic fields aligned with this axis (see  Fig.~2a in the main text). The black line depicts the magnetostriction calculated at zero temperature, using equations (\ref{2a}), (\ref{6a}) and (\ref{9a}) for the W1 and W2 electrons and assuming that $c_hB^2$ is the contribution of the holes to the magnetostriction. We use the following values of the parameters: $c_h=1.49\times 10^{-8}$ T$^{-2}$, $a_{W1}=-1.58\times 10^{-6}$, $a_{W2}=-0.6\times 10^{-6}$, $F_{W1}=7.2$ T, $F_{W2}=5$ T, $\gamma_{W1}=0.025$ and $\gamma_{W2}=0.1$.
} \end{figure}   

Finally, as an example, we calculate the magnetostriction produced by the two groups of the Weyl electrons (W1 and W2) and by the holes, assuming that the contribution of these holes to the magnetostriction are described by the term $c_hB^2$ where $c_h$ is a constant.
Let experimental data on the magnetostriction in the ultra-quantum regime for the W1 and W2 electrons can be approximated by the polynomial $a+bB+c_hB^2$. Then, if the frequencies $F_{W1}$, $F_{W2}$ are known, the constants $a_{W1}$ and $a_{W2}$ are found from the equations that follows from formula (\ref{11a}),
\begin{eqnarray*}
 a_{W1}+a_{W2}=a, \\
 -0.75\left(\frac{a_{W1}}{F_{W1}}+\frac{a_{W2}}{F_{W2}} \right)=b.
\end{eqnarray*}
With known $a_{W1}$, $a_{W2}$, $F_{W1}$, $F_{W2}$, the magnetostriction at an arbitrary $B$ is calculated, using formulas (\ref{2a}) and (\ref{6a}). The dimensionless Dingle temperatures are determined from the best fit of the calculated curve at $T=0$ to the appropriate experimental data at a low temperature. In particular,  we find for TaAs that $F_{W1}=7.2$ T, $a=-2.18\times 10^{-6}$, $b=2.55\times 10^{-7}$ T$^{-1}$, $c_h=1.49\times 10^{-8}$ T$^{-2}$. In the case of $F_{W2}=5$ T, the calculated magnetostriction together with the experimental data are presented in Supplementary Fig.~\ref{fig2a}.

\section{Supplementary Note $3$: Magnetostriction produced by electrons with parabolic spectrum}

For comparison, consider the magnetostriction $u_{zz}$ produced by the ``trivial'' electrons with the parabolic dispersion,
\begin{eqnarray}\label{17a}
\varepsilon(p_x,p_y,p_z)=\varepsilon_0 +\frac{p_x^2}{2m_x}+\frac{p_y^2}{2m_y}+\frac{p_z^2}{2m_z},
\end{eqnarray}
where $\varepsilon_0$ is the edge of the electron band, and $m_x$, $m_y$, $m_z$ are the effective electron masses. In the magnetic field $H$ these electrons have the well-known energy spectrum,
\begin{eqnarray}\label{18a}
\varepsilon_n(p_{\parallel})=\varepsilon_0+\frac{eB\hbar}{m_{*}}
(n+\frac{1}{2}+ \delta)+\frac{p_{\parallel}^2}{2m_{\parallel}},
\end{eqnarray}
where $n$ is an integer ($n\ge 0$);  $p_{\parallel}$ is the quasimomentum along the magnetic field; $m_{\parallel}$ and $m_{*}$ are the longitudinal effective mass and the cyclotron mass, respectively, and the electron magnetic moment composed of its spin and orbital parts is written as $\delta(e\hbar/m_*)$, with $\delta$ being a constant. Since due to the time reversal symmetry, the contribution of any charge-carrier pocket to the $\Omega$ potential for a noncetrosymmetric crystal  can be considered as a half contribution of the same pocket with the states that are doubly degenerate in spin, one should imply that the parameter $\delta$ in  the Zeeman term $\delta(e\hbar B/m_*)$ of Eq.~(\ref{18a}) takes the values $+\delta$ and $-\delta$ (in this situation, $\delta$ can be represented in terms of the $g$ factor, $\delta=gm_*/4m$). With this spectrum, we obtain at zero temperature,
\begin{eqnarray}\label{19a}
n(\zeta-\varepsilon_0,B)
=\frac{3n(\zeta- \varepsilon_0,0)}{4u^{3/2}} \!\! \sum_{\pm,n=0}^{n=[u-0.5\mp \delta]}\!\!\!\!\!\!\sqrt{u\!-\!\frac{1}{2}\!\mp\!\delta\!-\!n},~~~~
\end{eqnarray}
where $n(\zeta-\varepsilon_0,B)$ is the density of the quasiparticles in the magnetic field $H=B/\mu_0$ at a given chemical potential $\zeta$;  the frequency $F$ in the ratio $u\equiv F/B$ is still defined by formula (\ref{7a}). However, for the trivial electrons, the maximal cross sectional area $S_{\rm max}$ of their Fermi surface is proportional to $(\zeta-\varepsilon_0)$.

{\bf Ultra-quantum regime.}
At $\delta<1/2$ and $\mu_0H>B_{uq}=F/(0.5-\delta)$, the sum in equation (\ref{19a}) disappears. Then, with formula (\ref{2a}), we obtain  the constant  contribution of the trivial-electron pocket to the magnetostriction in this ultra-quantum limit,
\begin{eqnarray}\label{20a}
u_{zz}=-\Lambda n(\zeta- \varepsilon_0,0) \equiv a.
\end{eqnarray}
Interestingly, at $\delta=1/2$, this limiting value of $u_{zz}$ cannot be reached since at any $H$, the lowest Landau subband remains occupied by the electrons. In this situation Eq.~(\ref{19a}) reduces to formula (\ref{6a}) for the Weyl electrons.

When $\delta>1/2$, the behavior of the magnetostriction in the ultra-quantum limit cardinally changes. It follows from Eq.~(\ref{19a}) that
\begin{eqnarray}\label{21a}
n(\zeta\!-\!\varepsilon_0,B)\!\!&=&\!\!\frac{3n(\zeta- \varepsilon_0,0)}{4u^{3/2}}\sqrt{u\!-\!\frac{1}{2}+\delta}, \\
n(\zeta\!-\!\varepsilon_0,B)\!\!&=&\!\!\frac{3n(\zeta- \varepsilon_0,0)}{4u^{3/2}}\! \left[\!\sqrt{u\!-\!\frac{1}{2}+\delta}+\!\sqrt{u\!-\!\frac{3}{2} +\delta}\right]\!, \nonumber
\end{eqnarray}
for $1/2<\delta<3/2$ at $\mu_0H>F/(1.5-\delta)$ and for $3/2<\delta<5/2$ at $\mu_0H>F/(2.5-\delta)$, respectively. In both the cases, $n(\zeta-\varepsilon_0,B)\propto B^{3/2}$ at $B\gg F$.
However, if a limited range of the high magnetic fields is considered, formulas (\ref{21a}) lead to the curves $u_{zz}(B)$, the shapes of which are close to the straight line described by Eq.~(\ref{11a}) (cf.\ Fig.~1 in the main text).
Nevertheless, the slopes of these curves $\lambda\equiv du_{zz}/dB$ depend on $\delta$ and differ from the slope characteristic of the Weyl quasiparticles. Indeed, the slope for the Weyl fermions, $\lambda=b=0.75\Lambda n(\zeta-\varepsilon_d,0)/F$, is determined by the frequency $F$ only if $u_{zz}$ is measured in units of $\Lambda n(\zeta-\varepsilon_d,0)=-a$.

The special case $\delta=1/2$, for which the formula for the magnetostriction exactly coincides with the expression for the Weyl electron, separates the two essentially different behaviors of the magnetostriction in the ultra-quantum limit.

{\bf Weak magnetic fields ($\mu_0H\ll F$).}
At $\mu_0H\ll F$, the ratio $u$ in Eq.~(\ref{19a}) is large, and the sum contains many terms. Similarly to the case of the Weyl electrons, we find from formulas (\ref{2a}) and (\ref{19a}) that the magnetostriction $u_{zz}$ produced by the trivial electrons comprises the smooth and oscillating parts in the weak-field range, $u_{zz}=\bar u_{zz} +u_{zz}^{\rm osc}$. These smooth ($\bar u_{zz}$) and oscillating ($u_{zz}^{\rm osc}$) parts have the following forms:
\begin{eqnarray}\label{22a}
\bar u_{zz}&=&\Lambda n(\zeta-\varepsilon_0,0) \frac{3B^2}{8F^2}(\delta^2-\frac{1}{12})\equiv cB^2, \\
u_{zz}^{\rm osc}&=&\frac{3\Lambda n(\zeta-\varepsilon_0,0)}{ 4u^{3/2}}\sum_{\pm}\zeta(-1/2,\{u-0.5 \pm  \delta\}),~~~
\label{23a}
\end{eqnarray}
where $\{x\}\equiv x-[x]$ is fractional part of $x$.
Interestingly, at $\delta \to 0$ when the Zeeman energy is negligible as compared to the cyclotron spacing between the Landau subbands, we find that $c=-\Lambda n(\zeta-\varepsilon_0,0)/(32F^2)$. This value is of the opposite sign as compared to $c$ for the Weyl electrons. On the other hand, the phase of the oscillating part (\ref{23a}) at $\delta \to 0$ is shifted by $\pi$ with respect to the phase of the oscillations described by Eq.~(\ref{14a}). This shift is caused by the difference in the Berry phases for these two cases \cite{prl,m-sh19}. On the other hand, at $\delta=1/2$ the constant $c$ exactly coincides with that given by Eq.~(\ref{13a}), and equation (\ref{23a}) for the oscillating part  reduces to Eq.~(\ref{14a}).

Thus, we find that the $B$-dependence of $u_{zz}$ for the trivial electrons with $\delta\neq 1/2$ essentially  differs from the appropriate dependence for the Weyl quasiparticles (see also Fig.~1 in the main text). This result enables one to distinguish between the Weyl and trivial electrons with measurements of the magnetic-field dependences of the magnetostriction.  However, if $\delta=1/2$, the formulas for the magnetostriction exactly coincide in these two cases.

According to Ref.~\cite{g1,g2}, the quantity $\delta$ is close to $1/2$ for a band in a {\it centrosymmetric} crystal if this band is separated from another by a small gap that is less than the strength of the spin-orbit interaction. These bands are doubly degenerate in spin for such crystals. Strictly speaking, the spectrum (\ref{18a}) (with $\pm\delta$) is applicable to this situation only if the chemical potential measured from the edge of the band, $\zeta-\varepsilon_0$, is noticeably less than the gap $\Delta$ in the spectrum. However,  it turns out that even if $\zeta-\varepsilon_0\gtrsim \Delta$, the $B$ dependence of the magnetostriction for the gapped spectrum coincides with its $B$ dependence for Dirac electrons. In this context, it is worth noting that the spectrum of bismuth is sufficiently well described by a two-band model, the parameter $\delta$ is close to $1/2$ \cite{g2}, and so the field dependence of its magnetostriction \cite{Mi,Kuchler1} is reminiscent of the appropriate dependence for the Dirac electrons. However,  the situation essentially changes for the {\it noncentrosymmetric}  Weyl semimetals considered here. In particular, the holes in TaAs are located near the nodal ring that would occur in this semimetal  in neglect of the spin-orbit interaction \cite{Arnold}. The spin-orbit coupling lifts the degeneracy of the bands, and there are {\it four} close bands in the region of the Brillouin zone where the holes exist. In this situation, there is no reason to expect that $\delta\approx 1/2$ for them.

As is clear from Supplementary Note 1, if not only $\varepsilon_0$ but also the gap in the spectrum  $\Delta$ depend on a deformation (and $\zeta-\varepsilon_0$ is not-too-small as compared to this gap), the rigid-band approximation can fail. In this case, formula (\ref{2a}) will contain an additional term, and in the weak-field range, this term is proportional to $B^2$.

\subsection{Supplementary Note 4: Magnetostriction and magnetization}

Let us compare the magnetostriction with the magnetization,  which is considered as another thermodynamic probe of the Weyl electrons \cite{Moll,Zhang,modic}. The magnetization $M$ characterizes the change of the electron energy in the magnetic field, and this change occurs for all electrons, including those lying far below the Fermi level. In particular, a band completely filled by electrons can give a noticeable contribution $\chi_0H$ to the magnetization where $\chi_0$ is a constant. It is significant that the absolute value of this constant increases when a top of the filled band approaches the lowest boundary of unoccupied electron states at the same point of the Brillouin zone. This increase is due to virtual interband transitions of electrons in the magnetic field. On the other hand, the magnetostriction probes the energy of the interaction between electrons and the elastic deformations of the crystal, and within the rigid-band approximation, it is determined by the change of the charge-carrier density in the magnetic field. This density remains unchanged for the filled band, and so this band does not contribute to the magnetostriction.

In the weak magnetic fields, the magnetization of trivial electrons is proportional to $(\zeta-\varepsilon_0)^{1/2}B\propto n^{1/3}B$ \cite{abr} and increases with increasing $n$. In contrast, the coefficient $c$ in Eq.~(\ref{22a}), which characterizes the change of the electron density in the magnetic field, decreases with $n$,
  \[c \propto \frac{n}{F^2}\propto (\zeta-\varepsilon)^{-1/2}\propto n^{-1/3}.\]
As to a Weyl point, it was found (see review \cite{m-sh19} and the reference therein) that  the appropriate magnetization looks like  $M\propto \ln(|E_F-\varepsilon_d|)B \propto \ln(n^{1/3})B$. In other words, the smaller Weyl pocket, the larger contribution of the Weyl point to the $M$, and this increase is reminiscent of that given by  Eq.~(\ref{16a}) for the coefficient $c$:
 \[c\propto (\zeta-\varepsilon_d)^{-1}\propto n^{-1/3}.\]
However, these enhancements of the magnetization and of the magnetostriction have different origin. The magnetostriction of the Weyl electrons grows to higher values due to the decrease of their pocket, whereas in the case of the magnetization, its logarithmic enhancement is caused by the lower filled band since the Fermi level, i.e., the boundary of the unoccupied electron states, tends to the top of this band $\varepsilon_d$ when the Weyl pocket shrinks.

In the ultra-quantum limit, the magnetization of the Weyl electrons is proportional to $B\ln (CB/F)-6F$  where $C$ is a constant \cite{m-sh16,m-sh,m-sh19}, see also \cite{Moll,Zhang}. Except for the logarithmic factor, this formula looks like Eq.~(\ref{11a}). However, this factor has the same origin as the factor $\ln|E_F-\varepsilon_d|$ in the weak-field expression, and it can be obtained by the replacement of the difference $|E_F-\varepsilon_d|$ by the Landau-level spacing, which becomes larger than this difference in the ultra-quantum limit. As to the magnetization of the trivial electrons in this limit, it tends to zero \cite{Zhang} if the parameter $\delta$\,$<$\,1/2 and is proportional to $B^{3/2}$ in the opposite case of $\delta$\,$>$\,1/2. (Indeed, for $\delta$\,$>$\,1/2, simple estimates give: $n\propto B^{3/2}$, the $B$-dependent contribution to the $\Omega$ potential, $\delta\Omega\propto (\delta-1/2)(e\hbar B/m_*)n$, and $M=- \partial\delta\Omega/\partial B \propto B^{3/2}$.) By and large we may conclude that although the results for the magnetostriction and the magnetization are quite similar, they stem  from the different contributions to the thermodynamic potential and have different physical origin.

\section{Supplementary Note 5: Magnetostriction of ${\rm TaAs}$}\label{si5}

In Supplementary Notes $2$ and $3$ the $B$-dependences of the magnetostriction are analyzed under the assumption that a variation of the magnetic field does not change the chemical potential $\zeta$ of  electrons. This situation does can occur when a crystal contains a large  charge-carrier group that maintains the constancy of the chemical potential. However, in TaAs there are eight equivalent pockets of the W1 electrons, sixteen equivalent pockets of the W2 electrons, and eight equivalent pockets of the holes, with all the pockets being relatively small. In this case it is necessary to take into account the magnetic-field dependence of the chemical potential in analyzing the magnetostriction. This dependence $\zeta(B)$ is found from the conservation condition of the total density of the charge carriers,
\begin{eqnarray}\label{24a}
\sum_i[n_i(\zeta,B)-n_i(\zeta_0,0)]=0,
\end{eqnarray}
where the summation is over all the pockets, $\zeta_0$ is the initial value of $\zeta$ at $H=0$, and the charge-carrier density in the magnetic field for the $i$th pocket, $n_i(\zeta,B)$, is described by formula (\ref{6a}) for the Weyl electrons and by Eq.~(\ref{19a}) for the holes [or by some other expression if the spectrum of the holes differs from that given by Eq.~({\ref{18a})]. With the $B$-dependence of $\zeta$, formula (\ref{2a}) is rewritten as follows:
	\begin{eqnarray}\label{25a}
	u_{zz}&=&\sum_i^N\Lambda_i[n_i(\zeta,B)-n_i(\zeta_0,0)].
	\end{eqnarray}
	It is clear from Eqs.~(\ref{24a}) and (\ref{25a}) that if all the constants $\Lambda_i$ were equal, the magnetostriction $u_{zz}$ would be zero.
	
	Consider a special case when the magnetic field is considerably less than the fields $F_{i}/\mu_0$ for all the groups of the charge carriers. In this situation, equation (\ref{24a}) in the chemical potential $\zeta$ is approximately solvable since the difference $n_i(\zeta,B)-n_i(\zeta_0,0)$ can be rewritten as follows:
	\begin{eqnarray*}
		 n_i(\zeta,B)&-&n_i(\zeta_0,0)=n_i(\zeta,B)-n_i(\zeta_0,B)\nonumber \\
		&+&n_i(\zeta_0,B)-n_i(\zeta_0,0)
		\approx \nu_i(\zeta-\zeta_0)\nonumber \\
		&+&n_i(\zeta_0,B)-n_i(\zeta_0,0),
	\end{eqnarray*}
	where $\nu_i\equiv \partial n_i(\zeta_0,0)/\partial \zeta_0$, and  we have  taken into account that $\partial  n_i(\zeta,B)/\partial \zeta\approx \nu_i$ at the weak magnetic fields $\mu_0H\ll F_{i}$.
	Then, equations (\ref{24a}) and (\ref{25a}) give
	\begin{eqnarray}\label{26a}
	 (\zeta-\zeta_0)&=&\frac{\sum_i[n_i(\zeta_0,B)-n_i(\zeta_0,0)]}{\nu}, \nonumber \\
	u_{zz}&=&\sum_i \tilde\Lambda_i[n_i(\zeta_0,B)-n_i(\zeta_0,0)], \\
	\tilde\Lambda_i&=& \sum_{m}\frac{\nu_m}{\nu}(\Lambda_i-\Lambda_m),
	\nonumber
	\end{eqnarray}
	where $\nu\equiv  \sum_i \nu_i$. Therefore, in this weak-field range the $B$-dependence of the chemical potential can be taken into account by the renormalization of the constants $\Lambda_i$.
	For stronger magnetic fields, it is necessary to solve equation (\ref{24a}) numerically.
	
	In the case of TaAs, it is convenient to rewrite general formula (\ref{25a}) with the use of Eq.~(\ref{24a}) as follows:
	\begin{eqnarray}\label{27a}
	u_{zz}&=&(\Lambda_{W1}-\Lambda_h)\sum_{i=1}^8[n_i(\zeta,B)- n_i(\zeta_{0},0)] \nonumber \\ &+&(\Lambda_{W2}-\Lambda_h) \sum_{i=1}^{16}[n_i(\zeta,B)-n_i(\zeta_{0},0)],
	\end{eqnarray}
	where the summation is carried out over the W1 and W2 pockets only; the constants $\Lambda_{W1}$, $\Lambda_{W2}$, and $\Lambda_{h}$ refer to the W1, W2 electrons, and to the holes, respectively. This formula is valid at any strength of the magnetic fields. In the weak magnetic fields, $\mu_0H\ll F_{i}$, formulas (\ref{24a}) and (\ref{27a}) are equivalent to Eqs.~(\ref{26a}).
	
	{\bf Magnetic field parallel to the c axis.}
	When the magnetic field is parallel to the $c$ axis, all the pockets in the W1 electron group or in the W2 group or in the hole group  give identical contributions to the magnetostriction. Let  $F_{W1}$, $F_{W2}$, and $F_{h}$ denote the frequencies $F_i$ for the W1 and W2 electrons pockets and for the holes, respectively. According to Ref.~\cite{Arnold}, at this direction of $B$, one has $F_{W1}\sim 7$ T, $F_{W2}\sim 5$ T, and the field $F_{h}\sim 19$ T  for the holes is larger than the maximal field $16$ T in our experiments. Since  $F_{h}$ is large, we shall  consider the range $\mu_0H\le 16$ T as the low-field region for the holes and shall describe their contribution to Eq.~(\ref{24a}) as  follows:
	\begin{eqnarray}\label{28a}
	n_h(\zeta,B)&-&n_h(\zeta_0,0)=n_h(\zeta,B)-
	n_h(\zeta,0)\nonumber \\
	&+&n_h(\zeta,0)- n_h(\zeta_0,0)\nonumber \\
	\approx
	B^2\Big(\beta(\zeta_0)&+&\frac{d\beta(\zeta_0)}{d\zeta_0} (\zeta-\zeta_0)\Big)\!+\nu_h(\zeta_0)(\zeta-\zeta_0),~~~
	\end{eqnarray}
	where $\nu_h=\partial n_h(\zeta,0)/\partial \zeta$ is density of states for the holes in zero magnetic field, whereas the function $\beta(\zeta)$ defines the variation of the hole density for the low  magnetic fields,
	\[ n_h(\zeta,B)-n_h(\zeta,0)=\beta(\zeta)B^2. \]
	(The oscillation contribution to $n_h(\zeta,B)$ is neglected.)
	If the holes can be described by the parabolic spectrum (\ref{17a}), (\ref{18a}), one arrives at [compare with Eq.~(\ref{22a})],
	\begin{eqnarray}\label{29a}
	\beta(\zeta)=\frac{3n_h(\zeta,0)}{8F_{h}^2} (\delta^2-\frac{1}{12}), \nonumber \\
	\frac{d\beta(\zeta)}{d\zeta}=\frac{\beta(\zeta)}{2(\varepsilon_0 -\zeta)}, \\
	\nu_h(\zeta)=-\frac{3n_h(\zeta,0)}{2(\varepsilon_0-\zeta)}, \nonumber
	\end{eqnarray}
	where  $\varepsilon_0$ is the edge of the hole band, and we have taken into account that $n_h(\zeta,0)\propto -(\varepsilon_0-\zeta)^{3/2}$ and $n_h(\zeta,0)/F_{h}^2(\zeta)\propto -(\varepsilon_0-\zeta)^{-1/2}$ for the parabolic spectrum. However, we emphasize that formula (\ref{28a}) is applicable to the holes even though their spectrum is not described by the parabolic model, and their Fermi surface noticeably differs from an ellipsoid. The only requirement of the applicability is the condition of the low-field limit for them (i.e., $B<F_{h}$).
	
	Let us introduce the dimensionless deviation of the chemical potential from its initial value at $H=0$, $\delta\tilde \zeta \equiv (\zeta- \zeta_0 )/(\zeta_0 -\varepsilon_{d,W1})$ where $\varepsilon_{d,W1}$ is the energy of the Weyl points W1.
	Then, at $T=0$,  equation (\ref{24a}) that determines $\delta\tilde\zeta(B)$ is rewritten as follows:
	\begin{eqnarray}\label{30a}
	\!& &\!\!\left(\!1-\frac{3(1+\delta\tilde\zeta)}{4u_1} -\frac{3}{2u_1^{3/2}} \!\!\sum_{n=1}^{n=[u_1]}\!\!\sqrt{u_1(1+\delta\tilde\zeta)^2-n}\right) \nonumber \\
	\!&+&\!\frac{n_{W2}}{n_{W1}}\!\left(\!1\!- \!\frac{3(1+v\delta\tilde\zeta)}{4u_2} \!-\!\frac{3}{2u_2^{3/2}}\! \!\!\sum_{n=1}^{n=[u_2]}\!\!\!\!\sqrt{u_2(1+v\delta\tilde\zeta)^2 -n}\right)\nonumber \\
	\!&+&\!B^2(\tilde\beta_0+ \tilde\beta_1\delta\tilde\zeta)+ \tilde\nu_h\delta\tilde\zeta=0,
	\end{eqnarray}
	where $n_{W1}$, $n_{W2}$ are the densities of the electrons  at zero magnetic field (i.e., at $\zeta=\zeta_0)$; $v\equiv (\zeta_0-\varepsilon_{d,W1})/(\zeta_0-\varepsilon_{d,W2})$; $\varepsilon_{d,W2}$ is the energy of the Weyl points W2,
	\begin{eqnarray*}
		\tilde\beta_0 &\equiv& -\frac{\beta(\zeta_0)}{n_{W1}}, \\
		\tilde\beta_1 &\equiv& -\frac{(\zeta_0 -\varepsilon_{d,W1})}{n_{W1}}\frac{d\beta(\zeta_0)}{d\zeta_0}, \\
		\tilde\nu_h &\equiv& -\frac{(\zeta_0 -\varepsilon_{d,W1})}{n_{W1}}\nu_h(\zeta_0);
	\end{eqnarray*}
	$u_1=F_{W1}/B$, $u_2=F_{W2}/B$, and the frequencies $F_{W1}$, $F_{W2}$ are taken at $\zeta=\zeta_0$. The densities $n_i$ satisfies the relation $n_{W1}+n_{W2}-|n_{h}|=n_{\rm imp}$ where $n_h$ is the hole density at $H=0$, and the charge carrier density $n_{\rm imp}$ is caused by impurities. This $n_{\rm imp}$ specifies the doping in the sample. With  the doping,  $1+(n_{W2}/n_{W1})- (|n_h|/n_{W1})\neq  0$.

	In the case of the parabolic spectrum for the holes,
	we obtain from Eqs.~(\ref{29a}),
	\begin{eqnarray}\label{31a}
		\tilde\beta_0&=& \frac{3|n_h|}{(8F_{h}^2n_{W1})}(\delta^2-\frac{1}{12}),\nonumber \\
		\frac{\tilde\beta_1}{\tilde\beta_0}&=& \frac{(\zeta_0 -\varepsilon_{d,W1})}{2(\varepsilon_{0}- \zeta_0)},\\
		\tilde\nu_h&=& -\frac{3|n_h|(\zeta_0 -\varepsilon_{d,W1})}{2n_{W1}(\varepsilon_{0}- \zeta_0)}= -\frac{3|n_h|}{n_{W1}}\frac{\tilde\beta_1}{\tilde\beta_0}.\nonumber
	\end{eqnarray}
	Interestingly, if the magnetic field is tilted away from the $c$ axis, the quantities $F_{h}$, $\delta$, and hence $\tilde\beta_0$, will change, but the ratio $\tilde\beta_1/\tilde\beta_0$ will remain unchanged for the parabolic spectrum. Of course, the parameters  $\tilde\nu_h$, $\tilde\beta_0$, and $\tilde\beta_1$ may essentially differ from these estimates if the dispersion of the holes noticeably deviates from the parabolic law.

	According to expression (\ref{27a}), the general formula for the magnetostriction $\Delta L/L=u_{zz}$ of TaAs at zero temperature takes the form:
	\begin{eqnarray}\label{32a}
	& &\frac{\Delta L}{L}=A_{W1}\Big(1-\frac{3(1+\delta\tilde\zeta)}{4u_1} \nonumber \\ &-&\!\frac{3}{2u_1^{3/2}}\!\!\! \sum_{n=1}^{n=[u_1]}\!\!\!\sqrt{u_1(1+\delta\tilde\zeta)^2-n}\Big)\!
	+A_{W2}\Big(\!1-\frac{3(1+v\delta\tilde\zeta)}{4u_2} \nonumber \\
	&-&\!\frac{3}{2u_2^{3/2}}\!\!\! \sum_{n=1}^{n=[u_2]}\!\!\!\sqrt{u_2(1+v\delta\tilde\zeta)^2-n}\Big),
	\end{eqnarray}
	where  $A_{W1}\equiv -(\Lambda_{W1}-\Lambda_{h})n_{W1}$, $A_{W2}\equiv -(\Lambda_{W2}-\Lambda_{h})n_{W2}$, and $\delta\tilde\zeta(H)$ is found from Eq.~(\ref{30a}).
	
	To take into account the electron scattering by impurities, we use the representation (\ref{9a}) for the sums in Eqs.~(\ref{30a}) and (\ref{32a}) and replace $\zeta$ by $\zeta+i\Gamma_i$ where the quantity $\Gamma_i$ characterizes the scattering of charge carriers in the pocket $i$ ($i=$W1, W2); see the Supplementary Note $2$. This means that $\delta\tilde\zeta$ is replaced by $\delta\tilde\zeta+ i\gamma_{W1}$ or $\delta\tilde\zeta+i(\gamma_{W2}/v)$  in the appropriate terms of Eqs.~(\ref{30a}) and (\ref{32a}) where  $\gamma_i= \Gamma_i/(\zeta- \varepsilon_{Wi})$.

	The magnetostriction  at nonzero temperature can be calculated with formula (\ref{10a}). In order to apply this formula, let us introduce the new variable $x\equiv (\varepsilon- \zeta )/(\zeta_0 -\varepsilon_{d,W1})$ in integral (\ref{10a}) and the dimensionless temperature $t\equiv T/(\zeta_0 - \varepsilon_{d,W1})$. Then, Eq.~(\ref{10a}) is rearranged as follows:
	\begin{equation}\label{33a}
	u_{zz}(\zeta,T)=\frac{1}{4t}\int_{\!-\infty}^{\infty}\!\! dx  \frac{u_{zz}(x,0)}{\cosh^2(x/2t)},
	\end{equation}
	where $u_{zz}(x,0)$ is given by formula (\ref{32a}) in which the factors $(1+\delta\tilde\zeta)$ and $(1+v\delta\tilde\zeta)$ are replaced by $(1+\delta\tilde\zeta+x)$ and $(1+v\delta\tilde\zeta+vx)$, respectively.

	\section{Supplementary Note 6: Estimates of the parameters characterizing Fermi-surface pockets in ${\rm TaAs}$}

Parameters characterizing Weyl Fermi pockets can be estimated, using the data for the quantum-oscillation frequencies and for the appropriate cyclotron masses \cite{m-sh21a}. In particular, if the frequency $F$ produced by a Weyl pocket and the appropriate cyclotron mass $m_*$ have been measured at least for one direction of the magnetic field, the formula
	\begin{eqnarray}\label{34a}
	|\zeta - \varepsilon_d|=\frac{S_{\rm max}}{ \pi |m_*|}=\frac{2e\hbar F}{|m_*|},
	\end{eqnarray}
	enables one to find the position of the chemical potential $\zeta$ relative to the energy $\varepsilon_d$ of the Weyl point. Here $S_{\rm max}$ is the area of the extremal cross section perpendicular to the magnetic field for the Weyl pocket. The density $n_W$ of the Weyl charge carriers can be expressed in terms of directly-measurable frequencies of the quantum oscillations,
	\begin{eqnarray}\label{35a}
	n_W=\frac{N_W V}{(2\pi\hbar)^3},
	\end{eqnarray}
	where $V$ is the volume of a Weyl pocket in the Brillouin zone,
	\begin{eqnarray}\label{36a}
	V\!\!=\!\!\frac{4 [S_{\rm max}^{(1)}S_{\rm max}^{(2)}S_{\rm max}^{(3)}]^{1/2}}{3\pi^{1/2}}\!=\!\frac{8\sqrt{2}\pi (e\hbar)^{3/2}(F_1F_2F_3)^{1/2}}{3},~~~~
	\end{eqnarray}
	$N_W$ is the numbers of the equivalent Weyl pockets, and $F_i$ are the frequencies associated with the principal directions of the Fermi-surface ellipsoid. In other words,  $F_1$ and $F_3$ are the maximal and minimal frequencies produced by the pocket when the magnetic field rotates in various planes, and $F_2$ corresponds to the direction of $H$ perpendicular to the directions at which $F_1$ and $F_2$ occur. The cross sectional areas $S_{\rm max}^{(i)}$ correspond to the frequencies $F_i$, and these cross sections are mutually orthogonal.
	
	In the case of the parabolic dispersion of charge carriers, the formula that is similar to Eq.~(\ref{34a})  looks as follows:
	\begin{eqnarray}\label{37a}
	|\zeta - \varepsilon_d|=\frac{S_{\rm max}}{2 \pi |m_*|}=\frac{e\hbar F}{|m_*|},
	\end{eqnarray}
	and contains the additional factor $1/2$ as compared to Eq.~(\ref{34a}).

	Using the experimental data for the cross-sectional areas and the cyclotron masses (Table I in Ref.~\cite{Arnold}) and the calculated value $F_{W2}\sim 5$ T for $H$ parallel to the $c$ axis  \cite{Arnold}, we obtain with formulas (\ref{34a})-(\ref{36a}) that $n_{W1}\approx 2.5\times 10^{18}$ cm$^{-3}$, $\zeta_0- \varepsilon_{d,W1} \approx 28.4\pm 3.5$ meV \cite{m-sh21a}, and  $n_{W2}/n_{W1}\sim 0.15$, $\zeta_0-\varepsilon_{d,W2} \approx 11.9\pm 1$ meV. Since the oscillations in the magnetostriction of our sample and the quantum oscillations observed by Arnold et al. \cite{Arnold} for the W1 electrons have practically the same frequency at $H$ parallel to the $c$ axis, we conclude that  the Fermi-surface parameters estimated above are appropriate for our samples, too. For this reason, in the calculation with formulas of Supplementary Note $5$, we set
 \begin{equation}\label{38a}
v\equiv\frac{(\zeta_0-\varepsilon_{d,W1})}{(\zeta_0- \varepsilon_{d,W2})}\,=\,2.5, \ \ \  \frac{n_{W2}}{n_{W1}}\,=\,0.15.
 \end{equation}

Interestingly, in the case of the parabolic spectrum for the holes, we  similarly can find $\varepsilon_{0}-\zeta_0$ using Eq.~(\ref{37a}) and values of $F$ and $m_*$ measured for the hole at two directions of $H$ \cite{Arnold}. It turns that  $\varepsilon_{0}-\zeta_0$ lies in the interval $12.6$--$20.6$ meV if the data for $H$\,$\parallel$\,[100] are used, and $\varepsilon_{0}-\zeta_0$ is in the range $7.9$--$16.2$ meV if we use the data for $H$\,$\parallel$\,[110]. Therefore, it is reasonable to assume that $12.6\ {\rm meV} <\varepsilon_{0}-\zeta_0< 16.2$ meV. With this $\varepsilon_{0}- \zeta_0$ and the above $\zeta_0 -\varepsilon_{d,W1}$, we arrive at the estimate,
	\begin{eqnarray*}
		\frac{\tilde\beta_1}{\tilde\beta_0}= \frac{(\zeta_0 -\varepsilon_{d,W1})}{2(\varepsilon_{0}- \zeta_0)}\approx 0.8\div 1.3.
	\end{eqnarray*}
If the dispersion of the holes deviates from the parabolic dependence, the estimate of $\varepsilon_{0}-\zeta_0$  and the expression for $\tilde\beta_1/\tilde\beta_0$ can noticeably change. Below we do not  exclude the possibility of this deviation  since the hole pockets of the Fermi surface in TaAs are associated with the band-contact lines that would occur in absence of the spin-orbit interaction \cite{Arnold}.

	\section{Supplementary Note 7: Calculation of the magnetostriction along the $[001]$ direction and an analysis of the obtained parameters}

\begin{figure}[t] 
		\centering
		\vspace{+9 pt}
		\includegraphics[scale=1.]{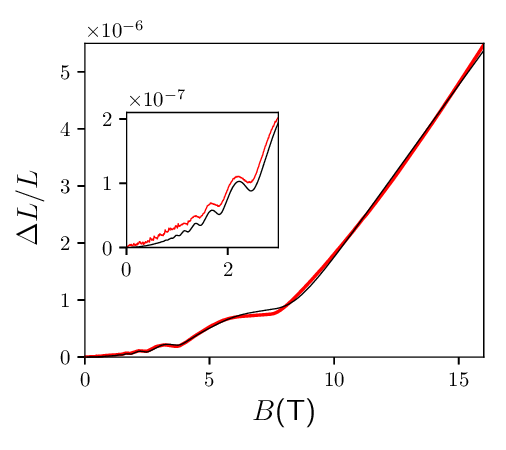}
		\caption{\label{fig3an}
Comparison of the calculated magnetostriction with the experimental data  for TaAs. 	The red line shows the $c$-axis magnetostriction $\Delta L/L$ of the crystal TaAs measured at $T=25$ mK and $H\parallel c$ (see  Fig.~2a in the main text). The black line depicts the magnetostriction calculated with equations (\ref{30a}), (\ref{32a}) [i.e. taking into account the $B$ dependence of $\zeta$] at zero temperature and at the values of the parameters presented in Supplementary Table I (set 1).
	} \end{figure}   

\textbf{Calculations.}
We calculate the magnetostriction with Eqs.~(\ref{30a}), (\ref{32a}) at given frequencies $F_{W1}=7.2$ T, $F_{W2}=5$, and at the fixed values of the ratios $(\zeta_0- \varepsilon_{d,W1})/(\zeta_0-\varepsilon_{d,W2})$ and  $n_{W2}/n_{W1}$, see Eq.~(\ref{38a}). The values of the other parameters are chosen so that the calculated magnetostriction best matches the experimental data (set 1 in Supplementary Table I). The result is presented in Supplementary Fig.~\ref{fig3an}.
Note that although the value $F_{W2}=5$ T was calculated in Ref.~\cite{Arnold}, the frequency $F_{W2}$ was not measured anywhere, and so it can, in principle, differ from this value. We find that if $F_{W2}$  decreases, the quality of such fits of the theoretical curve to the experimental data gradually improves, and the best fit is reached at $F_{W2}\approx 1.35$ T; see set 2 in Supplementary Table I and Supplementary Fig.~\ref{fig3a}. In this context, we shall analyze the two sets of the parameters below. The first set agrees with the value of $F_{W2}$ from Ref.~\cite{Arnold}, while the second one provides the best fit of the theoretical curve to the experimental data on the  magnetostriction.

\begin{table}
\caption{\textbf{The values of the parameters for the calculation of the magnetostriction $\Delta L/L$ along the $c$ axis and of the dependence $\zeta(B)$ at $H$ aligned with $c$ axis and at   $v\equiv (\zeta_0-\varepsilon_{d,W1})/(\zeta_0-\varepsilon_{d,W2})$\,=\,2.5, $n_{W2}/n_{W1}$\,=\,0.15.} }
\begin{tabular}{c|cccccccccc}
\hline
\hline \\[-2.5mm]
\#&$F_{W1}$&$F_{W2}$&$\tilde\nu_h$&$10^{3}\tilde\beta_0$&$10^3\tilde\beta_1$&$A_{W1}$ &$A_{W2}$&$\gamma_{W1}$&$\gamma_{W2}$ \\
 &T&T&~&T$^{-2}$&T$^{-2}$&$10^{-6}$&$10^{-6}$& &  \\
\colrule
1&$7.2$&$5$&$-0.91$&$1.63$&$1.53$&$-10.9$&$-1.92$&$0.025$&$0.1$ \\
2&$7.2$&$1.35$&$-1.39$&$5.93$&$1.34$&$-4.72$&$-0.43$&$0.025$&$0.1$
\\
\hline \hline
\end{tabular}
\end{table}

We also compare the magnetostriction measured at the temperature $T$\,=\,4.2\,K  with the magnetostriction calculated at a finite dimensionless temperature $t$\,=\,$T/(\zeta_0-\varepsilon_{W1})$, taking into account the dependence $\zeta(B)$. In Supplementary Fig.~\ref{fig5a}, we demonstrate this comparison in the case of the first set of the parameters in Supplementary Table I. The theoretical curve agrees with the experimental data at $t$\,$\approx$\,0.015. (The same $t\approx 0.015$ is also obtained for the second set.) This value of $t$ leads to the independent estimate of $\zeta_0-\varepsilon_{W1}$:  $\zeta_0-\varepsilon_{W1}$\,$\approx$ 280\,K\,$\approx$\,24\,meV, which is only a little less than the value 28.4\,$\pm$\,3.5\,meV obtained from the data of Arnold et al. \cite{Arnold} in Supplementary Note $6$.

We may now estimate the Dingle temperatures $T_{D,W1}=(\zeta_0- \varepsilon_{W1}) \gamma_{W1}/\pi$ for the W1 electrons. The data of Supplementary Table\,I and $\zeta_0-\varepsilon_{W1}\,\approx\,280$\,K give $T_{D,W1}$\,$\approx$\,2.2\,K. This value of  $T_{D,W1}$ is comparable with $T_{D,W1}$\,$\approx$\,3.2\,K obtained in Ref.~\cite{Arnold}. As to $T_{D,W2}$, we can only tentatively estimate its value  $T_{D,W2}=(\zeta_0- \varepsilon_{W2}) \gamma_{W2}/\pi \sim 3.6$\,K since the $F_{W2}$ oscillations are unobservable in experiments, and $\gamma_{W2}$ is not known  precisely. In our calculations, we have set $\gamma_{W2}=0.1$ only in order to suppress the oscillations with the frequency $F_{W2}$. Nevertheless, this value of $T_{D,W2}$ is reasonable. Assuming that the W1 and W2 electrons have approximately the same mean-free path $l_{\rm fp}$ and taking into account that $V_{W2}/V_{W1}\sim (16.52/1.92)^{1/3}\approx 2$ \cite{grassano} where $V_{Wi}$ are the mean velocities of the electrons  in these pockets, we obtain $T_{D,W2}\sim 4.4$ K from $l_{\rm fp}\sim (\hbar/T_{D,W1})V_{W1} \sim (\hbar/T_{D,W2})V_{W2}$. Note also that a reduction of $T_{D,W2}$ to 2.5 K only  marginally affects the agreement between the experimental and calculated data for the first set of the parameters and has no effect at all on the agreement for the second set when  $F_{W2}$ is small. Apart from  $T_{D,W2}>T_{D,W1}$, the smallness of $F_{W2}$ may be another reason why this frequency is not detected in experiments.

	\begin{figure}[t] 
		\centering
		\vspace{+9 pt}
		\includegraphics[scale=1.]{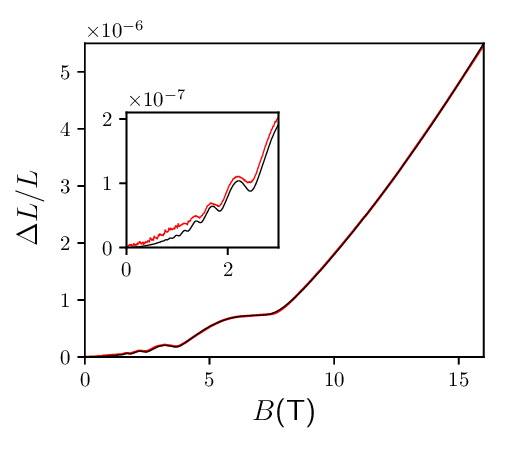}
		\caption{\label{fig3a} Comparison of the calculated magnetostriction with the experimental data  for TaAs.
			The red line shows the $c$-axis magnetostriction $\Delta L/L$ of the crystal TaAs measured at $T=25$ mK and $H\parallel c$  (see  Fig.~2a in the main text).
The figure is similar to Supplementary Fig.~\ref{fig3an}. In particular, the black line depicts the magnetostriction calculated with equations (\ref{30a}), (\ref{32a})  at zero temperature but with the other values of the parameters (set 2 in Supplementary Table I).
	} \end{figure}   

 \begin{figure}[tbp] 
 \centering
\includegraphics[scale=1]{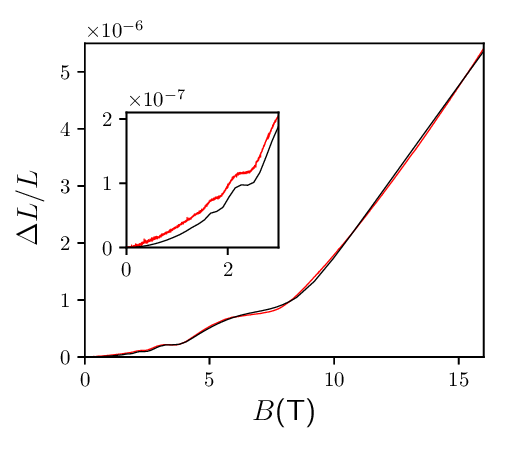}
\caption{\label{fig5a} Comparison of the calculated magnetostriction with the experimental data at $T$\,=\,4.2\,K for TaAs.
The red line shows the magnetostriction $\Delta L/L$ measured along the $c$ axis at $T$\,=\,4.2\,K and $H\parallel c$ (see  Fig.~2b in the main text). The black line corresponds to the magnetostriction calculated with Eqs.~(\ref{30a}), (\ref{32a}), (\ref{33a}), using the first set of the parameters in Supplementary Table I and the dimensionless temperature $t$\,=\,$T/(\zeta_0-\varepsilon_{W1})$\,=\,0.015. The inset is a zoom into the low-field region.
 } \end{figure}   

\textbf{Analysis of the results.}
Knowing $n_{W1}$, $n_{W2}/n_{W1}$, and the values of the parameters $a_{W1}=-\Lambda_{W1}^c n_{W1}$, $a_{W2}=-\Lambda_{W2}^c n_{W2}$ obtained in the neglect of the $B$ dependence of the chemical potential (see Supplementary Note 2 and Supplementary Table II), we can find the constants $\Lambda_{W1}^c$ and $\Lambda_{W2}^c$, Supplementary  Table II. Let us now estimate these constants in the case when the dependence of the chemical potential on $B$ is taken into account. The coefficient $c_h=1.49\times 10^{-8}$ found in the neglect of this dependence can be written as $c_h=-\Lambda_h^c n_{W1}\tilde\beta_0$ (see the definitions of $\tilde\beta_0$ and $\beta$ above). This relationship immediately gives $\Lambda_h^c$ with the data of Supplementary Table I and $n_{W1}\approx 2.5\times 10^{18}$. Then, the constants $\Lambda_{W1}^c$ and $\Lambda_{W2}^c$ can be calculated from the definitions $A_{W1}=-(\Lambda_{W1}^c-\Lambda_h^c)n_{W1}$ and $A_{W2}=-(\Lambda_{W2}^c-\Lambda_h^c)n_{W2}$.  The results of these calculations for both the sets of the parameters  are presented in Supplementary Table II. A comparison of the obtained constants $\Lambda_i^c$ shows that in describing the magnetostriction, the use of the simplified approach, for which the $B$ dependence of $\zeta$ is neglected, is reasonably well justified. (A caution should be given only to the case when one the parameters $a_{Wi}$ is unusually small.) The same conclusion follows either from from a comparison of Supplementary Figs.~\ref{fig2a} and \ref{fig3an} or from the calculation of the  dependence $\zeta(B)$ in the weak magnetic fields. Formula (\ref{30a}) for such fields yields $\tilde\delta\zeta\approx zB^2$ where the coefficient $z$ is determined from the equation,
	\begin{eqnarray*}
(3+3v\frac{n_{W2}}{n_{W1}}-\tilde\nu_h)z= -\frac{1}{16F_{W1}^2}+\frac{n_{W2}}{n_{W1}}\frac{1}{16F_{W2}^2} -\tilde\beta_0.
	\end{eqnarray*}
With the data of Supplementary Table I, we find that this coefficient is equal to $z_1\approx 10^{-5}$ for the first set and to $z_2\approx -7.6\times 10^{-5}$ for the second one. On the other hand, in the weak-field range, formula (\ref{32a}) for the magnetostriction reduces to the expression,
	\begin{eqnarray*}
\frac{\Delta L}{L}=-A_{W1}\!\left(3\delta\tilde\zeta +\!\frac{B^2}{16F_{W1}^2}\right)\!-\!A_{W2}\!\left(3v\delta\tilde\zeta +\!\frac{B^2}{16F_{W2}^2}\right).
	\end{eqnarray*}
Comparing, e.g., the terms $3\delta\tilde\zeta$ and  $B^2/(16F_{W1}^2)$, one obtains that the first term is smaller than the second one by the factor $40$ for set 1 and by the factor $5$ for set 2. In other words, in the first approximation, one really can neglect the dependence $\zeta(B)$.

\begin{table}
\caption{\textbf{The values of $a_{Wi}$ and $\Lambda_{i}^c$ obtained with and without the neglect of the dependence $\zeta(B)$.}}
\begin{tabular}{c|cccccc}
\hline
\hline \\[-2.5mm]
set&$\zeta=\zeta(B)$&$\Lambda_{W1}^c$&$\Lambda_{W2}^c$&$\Lambda_{h}^c$&
$a_{W1}$&$a_{W2}$ \\
 \#&~&$10^{-24}$cm$^{3}$&$10^{-24}$cm$^{3}$&$10^{-24}$cm$^{3}$&
 $10^{-6}$&$10^{-6}$  \\
\colrule
$1$&$-$&$0.63$&$1.6$&~&$-1.58$&$-0.60$ \\
$1$&$+$&$0.71$&$1.47$&$-3.7$&~&~
\\ \colrule
$2$&$-$&$0.85$&$0.17$&~&$-2.12$&$-0.062$ \\
$2$&$+$&$0.89$&$1.42$&$-1.0$&~&~ \\
\hline \hline
\end{tabular}
\end{table}

Assuming the parabolic spectrum of the holes and inserting the data of Supplementary Table I for set 1 into formulas  (\ref{31a}), we  obtain that $\tilde\beta_1/\tilde\beta_0 \approx 0.94$ and  $|n_h|/n_{W1}$\,$\approx$\,0.32. The ratio $\tilde\beta_1/\tilde\beta_0$ agrees with the estimate derived from the data of Ref.~\cite{Arnold} (Supplementary Note $6$). The obtained value of $|n_h|/n_{W1}$ means that the doping in our sample, $n_{W1}+n_{W2}-|n_h|$, is of the order of $0.83n_{W1}\,\approx\,2\times 10^{18}$\,cm$^{-3}$. As to the value of $\tilde\beta_0$ presented in Supplementary Table\,I  for set 1, it can be reproduced with the first formula of (\ref{31a}) at $|n_h|/n_{W1}$\,=\,0.32 and $F_{h}$\,=\,19\,T  if we assume that due to the spin-orbit interaction, the magnetic moment of the holes is sufficiently large  (the parameter $\delta$ amounts to $2.2$).
	
A similar analysis for the second set of the parameters in Supplementary Table I leads to the unrealistic value of $\varepsilon_0-\zeta_0 \approx 63$  meV which essentially exceeds the estimate  $\varepsilon_0-\zeta_0$\,$\approx$\,12.6$\div$16.2\,meV  inferred  from the results of Arnold et al.\ \cite{Arnold} under the assumption of the parabolic spectrum for the holes (Supplementary Note $6$). Moreover, formulas (\ref{31a}) with set 2 give $|n_h|/n_{W1}$\,$\approx$\,2.05, the value of which  disagrees with the electron doping of the specimen that was used in Ref.~\cite{Arnold} and that is similar to ours. These discrepancies indicate that the second set of the parameters may be admissible only if a significant deviation of the  dispersion of the holes from the parabolic law occurs and hence if formulas (\ref{31a}) and the estimate for $\varepsilon_0-\zeta_0$ in Supplementary Note $6$ are not applicable to this charge carriers.

	\section{Supplementary Note 8: Magnetostriction along the a axis}

Consider the magnetostriction $\Delta L/L$ along the $a$ axis for the magnetic field still aligned with the direction $[001]$. In this case, in describing the magnetostriction, only the values of $A_{W1}$ and $A_{W2}$ can change due to a change of the constants $\Lambda_{W1}$, $\Lambda_{W2}$, and $\Lambda_{h}$. All the other parameters determining this $\Delta L/L$ should remain the same as in the case of the $c$-axis magnetostriction. Therefore, a comparison of theoretical and experimental results for the $a$-axis magnetostriction may enable one to choose between the two sets of the parameters presented in Supplementary Table I.

\begin{figure}[t]
	\begin{center}
\includegraphics[scale=1.00]{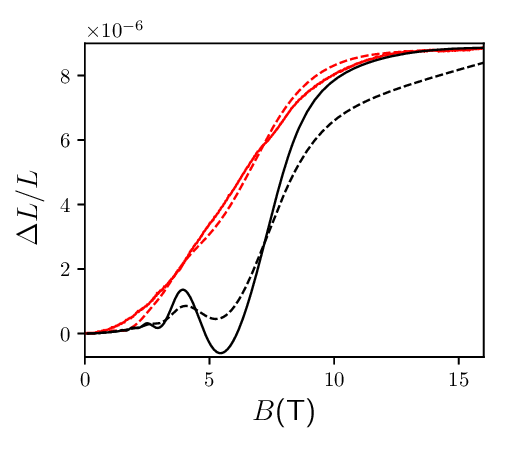}
\caption{\label{fig6a} Comparison of the calculated magnetostriction with the experimental data for TaAs. The solid red  line shows the relative length change $\Delta L/L$ of TaAs (sample 2) measured along the [100] direction at 25\,mK for the magnetic field aligned with the $c$ axis. The dashed red line depicts the $a$-axis magnetostriction calculated at $T$\,=\,0, using set 2 of the parameters in Supplementary Table I, but with new $A_{W1}=15.7\times 10^{-6}$, $A_{W2}=-2.61\times 10^{-6}$, $\gamma_{W1}=0.1$, and  $\gamma_{W2}=0.2$. The solid black line corresponds to set 1, but with $A_{W1}=1.48\times 10^{-6}$, $A_{W2}=-17.2\times 10^{-6}$, $\gamma_{W1}=0.06$, and  $\gamma_{W2}=0.1$. The dashed black line is plotted for the same parameters as the solid black curve except for $\gamma_{W1}=0.1$, and  $\gamma_{W2}=0.2$.
}	
\end{center}
\end{figure}

The magnetostriction $\Delta L/L$ measured along the $a$ axis at $B$ parallel to $[001]$ is shown in Supplementary Fig.~\ref{fig6a}.
In this figure, we also show the magnetostriction calculated with both sets of the parameters in Supplementary Table I, varying only the constants $A_{W1}$ and $A_{W2}$. (We also use modified values of $\gamma_{W1}$ and  $\gamma_{W2}$ in order to suppress the oscillations; note that Supplementary Figs.~\ref{fig3an}-\ref{fig5a} and \ref{fig6a} show the magnetostrictions for the two  different samples). For the first set, we are not able to match well the theoretical curve with the experimental data. The solid black line shows the best result that we have been able to obtain. An increase of $\gamma_{Wi}$ even enhances the disagreement (the dashed black line), and a subsequent variation of $A_{Wi}$ does not improve this situation noticeably. For the second set, the obtained  theoretical curve approximately reproduces these data if we take $A_{W1}$\,=\,15.7\,$\times$\,10$^{-6}$ and $A_{W2}$\,=\,$-$2.61\,$\times$\,10$^{-6}$ together with the modified values of $\gamma_{W1}$\,=\,0.1 and  $\gamma_{W2}$\,=0.2. Since the parameter $\lambda\equiv (1/L)d\Delta L/dB$ in the ultra-quantum region $B\gtrsim 8$ T has the form $2c_hB +b$, the slope of this linear dependence gives the parameter $2c_h$ (see Supplementary Fig.~\ref{fig:Fig1}). Similarly to Supplementary Note 7, with the obtained  $A_{W1}$, $A_{W2}$, $c_h\approx -6.25\times 10^{-8}$, and with set 2 from Supplementary Table I, we find the constants $\Lambda_{W1}^{\perp}$, $\Lambda_{W2}^{\perp}$, and $\Lambda_{h}^{\perp}$ defined in Supplementary Note $1$ (Supplementary Table III). Knowing these constants $\Lambda_i^{\perp}$ and the constants  $\Lambda_i^c$ (set 2 in Supplementary Table II), one can calculate  $\lambda_{zz}^{(i)}$ and $\bar\lambda_{xx}^{(i)}$ from Eqs.~(\ref{5a}); Supplementary Table III. Note that the values of $\lambda_{zz}^{(i)}$ and $\bar\lambda_{xx}^{(i)}$ make it possible to predict how the frequencies of quantum oscillations in magnetic fields for the W1 and W2 electrons and for the holes change under a uniform compression. It is clear from the data of Supplementary Table III that the Fermi surface of the W2 electrons should be most sensitive to this compression.

\begin{table}
\caption{\textbf{The values of $\Lambda_i^c$, $\Lambda_i^{\perp}$, $\lambda_{zz}^{(i)}$, and $\bar\lambda_{xx}^{(i)}$ for the second set of the parameters in Supplementary Table I (i=W1, W2, h).}}
\begin{tabular}{c|cccc}
\hline
\hline \\[-2.5mm]
~&$\Lambda_i^c$&$\Lambda_i^{\perp}$&$\lambda_{zz}^{(i)}$&$\bar\lambda_{xx}^{(i)}$ \\
~&$10^{-24}$cm$^{3}$&$10^{-24}$cm$^{3}$&eV&eV \\
 \colrule
W1&$0.89$&$-2.1$&$2.2$&$5.7$ \\
W2&$1.42$&$11.2$&$-21.7$&$-35.5$ \\
h&$-1.0$&$4.2$&$-5.6$&$-12$ \\
\hline \hline
\end{tabular}
\end{table}

The presented fit of the theoretical curve to the experimental data   argues in favor of set 2. However, we emphasize that the orientation of the sample in respect to the magnetic field direction cannot be established in our experiments with a high accuracy, and hence the experimental curve presented in Supplementary Fig.~\ref{fig6a} does not necessarily corresponds to the case of $B$\,$\parallel$\,[001]. Due to the high sensitivity of the $a$-axis magnetostriction to the angle between $B$ and the $c$ axis (see Fig.~6 in the main text),  its true $B$-dependence at $B$\,$\parallel$\,[001] may essentially differ from the experimental curve presented in Supplementary Fig.~\ref{fig6a}, and so a more precise orientation of the sample is required in order to reliably exclude the possibility of set 1.

Let us now discuss the unusual high sensitivity of the $a$-axis magnetostriction to small deviation of $B$ from the $c$ axis (Fig.~6 in the main text). When the magnetic field is tilted away from the direction $[001]$ in the plane (010), electron pockets in the W1, W2 groups, as well as the hole pockets, produce different contributions to the magnetostriction, and one should take into account that with increasing the tilt, the parameters $F_{W1}$, $F_{W2}$, $F_{h}$ not only change in magnitude but also the number of these parameters increases (the charge-carrier pockets in any group have different values of these parameters). This makes the strict theoretical analysis of the magnetostriction very complicated. However, the low sensitivity of the $c$-axis magnetostriction to the deviation angle (Fig.~7 in the main text) indicates that the total differences of the densities $n(B)-n(0)$ for each of the groups (W1, W2, h) change only slightly with the angle. As is explained in Supplementary Note 1, the $a$-axis magnetostriction contains an additional term that is proportional to $n_{a1}(B)-n_{a2}(B)$ where $n_{a1}$ and $n_{a2}$ are the charge-carrier densities of the equivalent pockets lying near the reflection planes (010) and (100), respectively. Consider the situation when equivalent pockets  have the shape of elongated ellipsoids, the longest axes of which are practically perpendicular to the $c$ axis. This model seems to be applicable to the Fermi surface of the holes in TaAs. When the magnetic field is tilted  in the (010) plane, the appropriate extremal cross-sectional areas of the ellipsoids lying near this plane abruptly decreases, whereas the appropriate areas for the ellipsoids located near the (100) plane remain practically unchanged. This means that in the weak magnetic fields, the difference $|n_{a1}(B)-n_{a2}(B)|$ steeply increases with the tilt angle. Calculating $\lambda(B)\equiv (1/L)d\Delta L/dB$ for the curves shown in Fig.~6 of the main text and approximating these $\lambda(B)$ by $2c_hB+b$ at $B>8$ T, we find the following values of  $c_h(\Theta)$ (in units of $10^{-8}$ T$^{-2}$):
  \begin{eqnarray*}
c_h(90^{\circ})&\approx& -6.25,\ \ \ c_h(87^{\circ})\approx 0, \\ c_h(86^{\circ})&\approx& 4.5,\ \ \ c_h(81^{\circ})\approx 3.8,
  \end{eqnarray*}
which do indicate the large variation of $c_h$ with the angle $\Theta$ between $B$ and the $a$ axis. This sharp variation may be due to the above-mentioned strong $\Theta$ dependence of  $|n_{a1}(B)-n_{a2}(B)|$ for the holes.  We cannot also exclude that the Fermi-surface pockets for the W2 electrons have a shape other than spherical, and so they can contribute to the angular dependence of the $a$-axis magnetostriction as well. (Although this assumption does not agree with the numerical calculation  of the Fermi surface for the W2 electrons \cite{Arnold}, however, subtle details of this small surface  can hardly be obtained in the calculations.) On the other hand, the W1 pockets have the shape close to ellipsoids elongated in the [001] direction, and so they give a gradually varying contribution to this angular dependence.

\end{document}